%% file: manuscript.tex
\documentclass[preprint,12pt]{elsarticle}

\usepackage{amssymb}
\usepackage{amsmath}

\input{preamble}

\input{notation}

\newcommand{\bogus}[1]{}

\journal{arXiv}

\begin{document}

\begin{frontmatter}



\title{Revealing Low-Dimensional Structure in 2D Richtmyer-Meshkov Instabilities via Parametric Reduced-Order Modeling}


\author{Daniel Messenger, Daniel Serino, Balu Nadiga, Marc Klasky} 

\affiliation{organization={Los Alamos National Laboratory},
            city={Los Alamos},
            postcode={87544}, 
            state={New Mexico},
            country={USA}}

\begin{abstract}
Efficient modeling of the Richtmyer-Meshkov instability (RMI) is essential to many engineering tasks, including high-speed combustion and drive and capsule geometry optimization in Inertial Confinement Fusion (ICF). In the latter, RMI causes the ablator and fuel to mix, introducing cold spots into the fuel and lowering performance; controlling RMI is thus a core ICF design concern. In this work, we introduce a reduced-order model for two-dimensional RMI based on the Latent Space Dynamics Identification (LaSDI) algorithm. We demonstrate the efficacy of the proposed methodology in efficiently parametrizing the solution space over a high-dimensional parameter vector consisting of material EOS parameters and initial conditions known to affect RMI growth rates. Using only late-time partial observations of the dynamics, we use our framework to not only provide a highly efficient dynamic surrogate model, but to reveal that the RMI exhibits the structure of a surprisingly low-dimensional and linear dynamical system, into the nonlinear growth regime, after a suitable nonlinear transformation is applied to the material interface, which we approximate as a trained autoencoder. Our use of practical observables and fundamental parameters suggests that such ROMs may be useful for downstream engineering tasks which confront the RMI, while the low-dimensional representation suggests a new direction for theoretical work.
\end{abstract}


\begin{highlights}
\item We present a parametric reduced-order modeling (ROM) framework for the Richtmyer-Meshkov Instabililty (RMI) in 2D geometries
\item The ROM is highly effective at capturing the evolution of material interfaces by directly modeling a material indicator function
\item ROM solutions generalize over a high-dimensional parameter space consisting of material EOS parameters and initial conditions
\item The proposed framework reveals that 2D RMI  exhibits a counterintuitive representation as a low-dimensional linear dynamical system 
\end{highlights}

\begin{keyword}
Richtmyer-Meshkov instability \sep  reduced-order model \sep  parametric PDEs \sep  Latent-Space Dynamics Identification (LaSDI)



\end{keyword}

\end{frontmatter}



\section{Introduction}
Hydrodynamic instabilities have been the subject of fervent study for decades due to the adverse (or in some cases, beneficial) effects they have in engineering applications~\cite{spakovszky2023instabilities,dimotakis2005turbulent}. They pose significant challenges to predictive modeling in inertial confinement fusion (ICF), where a fusion reaction is initiated by driving a series of shells inward to compress a target fuel source. In ICF, shocks are generated which cross material interfaces (e.g. ablator $\leftrightarrow$ DT-ice and DT-ice $\leftrightarrow$ DT-gas), depositing baroclinic vorticity and launching the Richtmyer–Meshkov instability (RMI) if the interface is even slightly perturbed \cite{aglitskiy2010basic,velikovich2000richtmyer,goncharov2009ablative}. Interface perturbations initially grow linearly under RMI, then later transition to nonlinear growth which leads to mixing that can degrade compression and yield \cite{remington1992large,marinak2024numerical}. The RMI is known to seed the more general Rayleigh-Taylor instability (RTI) in ICF capsules (the RMI is a special case of RTI in which the incident wave is supersonic, i.e.\ a shock wave), where the RTI is known to initially grow exponentially, hence RTI seeding can further accelerate mixing \cite{aglitskiy2010basic}. 

Like the RTI, the RMI is a ubiquitous hydrodynamic instability across many scientific disciplines (see \cite{zhou2021rayleigh} and references therein). As discussed, the RMI in ICF severely degrades efficiency of energy deposition onto the fuel source and subsequent fuel ignition by causing fuel to mix \cite{kilkenny1994review}. In supernovae, the RMI breaks spherical symmetry in the expansion of stellar gases, resulting in the formation of heavy mass elements \cite{abarzhi2019supernova}. The RMI also enhances supersonic mixing in shock-flame interactions \cite{ciccarelli2008flame}. In contrast to RTI, the RMI presents specific challenges to experimental measurement, computational modeling, and design optimization due to its shock-driven onset. The supersonic nature of interface acceleration under RMI brings into question the dependence of RMI evolution on the sound speeds and Hugoniot relations of materials involved, which fundamentally links the dynamics to the material equations of state.

The canonical setting for theoretical study of RMI consists of a planar interface between two materials and an incident planar shock. {\it Modal} analysis has been utilized extensively to model the RMI in this setting, where the material interface perturbation is decomposed into Fourier modes. Simplified analytical models exist for the dynamics of the peak-to-trough distance of individuals modes, where it is well-known that individual modes first exhibit uncoupled linear growth immediately after the shock passage \cite{zhou2021rayleigh}, followed by weakly nonlinear growth in which peak-to-trough dynamics are mode-coupled \cite{haan1991weakly}. 
While single-mode growth rate models in the weakly nonlinear phase have seen much in the way of experimental validation \cite{holmes1999richtmyer}, modifications continue to be developed to match experiments in more realistic settings, such as quasi-single-mode interfaces in compressible materials \cite{liang2019richtmyer}. Modeling the strongly nonlinear phase of RMI growth is an active area of research, where modal analysis is not applicable due to the multi-valued nature of the interface curve and typically full-order hydrodynamics simulations must be relied on \cite{thornber2017late,zhou2020mode}. Reliable and efficient reduced-order models (ROMs) for RMI in practical settings are thus largely unavailable.

The current article is motivated by this growing need for efficient ROMs for multi-mode RMI into the nonlinear growth phase as needed in outer-loop applications (e.g.\ parameter inference and design optimization in ICF \cite{maltba2025causal}), where classical simulation methods are typically infeasible due to the non-differentiability of first-principles physics codes\footnote{We note that several recent works have also presented differentiable hydro codes \cite{bezgin2023jax,lim2023fully} to accelerate inverse problems. Practical differentiable multiphysics codes are as-of-yet unavailable \cite{bramwell2025differentiable}.}. To isolate the problem of parametrized RMI onset and evolution, we restrict our focus to a simplified hydrodynamic setting, leaving a realistic multiphysics treatment of RMI as occurs in ICF to future work. Still, this represents a significant challenge from the parametric modeling perspective, as even the linear RMI growth phase is subject to a high-dimensional parameter vector over which to optimize, as linear growth rates are approximately given by $\dot{a} \approx k (A^+\Delta u )a_0^+$, where $a(t)$ is the interface amplitude, $a_0^+$ is the initial post-shock amplitude, $k$ is the dominant wavenumber of the interface, $A$ is the Atwood number, and $\Delta u$ is the shock speed. Thus, even in the simplest setting, the RMI growth rate depends on material parameters, interface geometry, and shock speed\footnote{The authors of \cite{weber2009richtmyer} and \cite{jiang2016parameterization} perform parametric studies of RMI in experimental and computational settings, respectively, without discussing a parametrized ROM for the RMI dynamics.}. 

Progress in the direction of design optimization under RMI was made in \cite{sterbentz2022design}, where the authors consider the problem of suppressing RMI growth in 2D planar geometries using gradient-free optimization, however their approach requires full hydrodynamic simulations during the inner loop and {\it a priori} knowledge of bubble and spike locations. The authors of \cite{sterbentz2024explosively} minimize RMI jet penetration using a neural network surrogate model for the objective function, which does not require in-the-loop hydro simulations. Another series of recent works have attacked RMI-related inverse problems, utilizing attention-based deep learning to reconstruct material interfaces and estimate parameters under RMI growth using the outgoing shock harmonics extracted from noisy (synthetic) radiographic images  \cite{huang2022physics,hossain2022high,serino2024reconstructing,gautam2025learning}. Image deblurring in the context of density reconstruction under hydrodynamic instabilities was also performed in \cite{stamlerclariphy}. 

The current work is distinct from these previous studies in that we present a {\it dynamic} model for the material interface under RMI, rather than a static image- or feature-inversion model, or a surrogate for a static objective function. While a direct surrogate as in \cite{sterbentz2024explosively} is possible in specific settings, a dynamic surrogate enables one to rapidly query the dynamics and retrieve the material interface over a range of input parameters and points in spacetime. To the best of the authors' knowledge, very few works in the literature present such dynamic, data-driven, parametric ROMs for hydrodynamic instabilities. A related work to ours by Cheung et al \cite{cheung2023local} develops a ROM framework for the Rayleigh-Taylor instability using patch-based Galerkin POD ROMs, which is distinct from the present RMI problem setting (e.g.\ in the geometry, observables, and degree of parametrization), in addition our ROM methodology is non-intrusive with respect to the full-order dynamics. Our conclusions also have different physical implications from \cite{cheung2023local}, namely we find that the RMI exhibits a low-dimensional structure.

Our proposed data-driven framework is capable of predicting RMI evolution into the nonlinear growth phase parametrized over relevant physical quantities. We use synthetic observables relevant to ICF experiments, that is, we consider observational data consisting of only the material interface at limited snapshots before and after shock passage, motivated by recent work demonstrating viability of extracting the material interface from noisy radiographic images \cite{serino2024reconstructing,gautam2025learning}. This leaves many solution compartments unobserved, including velocity and temperature fields, as well as the full material density, consistent with the fact that direct observation of state variables in the ICF setting is not possible. 
The ROM consists of a parametrized dynamical system for the evolution of the material interface via autoencoder compression and linear time evolution in a low-dimensional latent space, parametrized by an auxiliary parameters-to-coefficients network incorporating material parameters and initial conditions. As such, it can be seen as an application of and extension to the Latent Space Dynamics Identification (LaSDI) framework \cite{fries2022lasdi} (for a review of related parametrized ROMs, see \cite{farenga2025latent,bonneville2024comprehensive}). Previous latent space dynamics methods have focused on fully observed high-order dynamics and low-dimensional parameter spaces. In contrast, our proposed framework utilizes only observations at the material interface and provides surrogate modeling capabilities for parametrized time evolution of the RMI in a relatively high-dimensional parameter space. We also utilize weight initialization at POD modes, which not only accelerates training but leads to more accurate ROMs (see \ref{app:PODvsrandom}). We find that previous methods used to parametrize the solution space in LaSDI models under low-dimensional parameter vectors are not applicable here, such as Gaussian process interpolation \cite{bonneville2024gplasdi}, radial basis functions \cite{fries2022lasdi}, and bivariate splines \cite{fries2022lasdi}; we show instead that a neural-network-based parameters-to-coefficient map can successfully parametrize the solution over the high-dimensional parameter space considered.

In addition to presenting a parametrized RMI modeling framework that is efficient and practical from an engineering perspective, the present work may be the first to demonstrate that evolution of a 2D material interface under RMI in convergent geometries, into the nonlinear growth phase, is intrinsically {\it low-dimensional}, in particular the intrinsic dimension is lower than the number of active Fourier modes at the material interface, and reduces to {\it linear time evolution}, after nonlinear mapping to the latent space. To the best of the authors' knowledge, these findings are not present in the literature, have not been derived previously, and are not revealed by other dimensionality estimation methods (e.g.\ singular value decay, local PCA). This indicates that LaSDI-type methods may be useful in revealing the dimensionality of dynamic physical phenomena, where other methods are impractical. We hope this observation will motivate future derivations of simplified models for nonlinear-phase RMI.

\subsection{Paper organization}\label{sec:paperorg}
The manuscript is organized as follows. In Section \ref{sec:problemsetting} we present the problem setting, including the full-order model, problem geometry, equation-of-state, parametric dependencies, and simulated observational setting in which we develop our ROMs. In Section \ref{sec:existingmodels} we also review existing reduced RMI modeling results from the literature, broadly covering results from theory, experiment, and scientific computing. In Section \ref{sec:methodology} we describe the proposed ROM framework, including compression and decompression operators, latent space dynamics, parameters-to-coefficients map, and optimization algorithms. In Section \ref{sec:results} we present our main results, from which we conclude that our ROMs serve as accurate forward models for the material interface under RMI, and that the optimal latent space dimension remains three under successively increasing parameter scopes. 
Section \ref{sec:discussion} contains a discussion and outlook for future research. 

\section{Problem setting}\label{sec:problemsetting}
We now describe the specific problem setting we consider for parametrized model-order reduction of RMI dynamics in 2D convergent geometries. This includes the full-order dynamics (Section \ref{sec:FOdyn}), as well as physical assumptions on parameters space and the nature of the synthetic ``observed'' data (Section \ref{sec:data}). We also include in Section \ref{sec:existingmodels} an overview of related research on RMI, from which we argue that the results presented here may be useful to broad audience, having both computational and theoretical implications.  

\subsection{Full-order dynamics}\label{sec:FOdyn}
The evolution of a material interface over time under RMI, as occurs in a dynamic ICF experiment, can be modeled by a system of partial differential equations (PDEs) describing radiation hydrodynamics. To isolate the RMI evolution, we utilize the compressible Euler equations in a manner analogous to 2D ICF-like benchmark problems in \cite{galera2010two,bello2020matrix}, leaving ablative RMI and radiation effects to future work \cite{aglitskiy2010basic}. The full-order model we consider is thus
\begin{subequations}\label{eq:euler}
\begin{align}
    \partial_t \rho +\nabla \cdot (\rho\ubf) &= 0 \\ 
    \partial_t (\rho \ubf) +\nabla \cdot (\rho\ubf \otimes \ubf + p\Ibf) &= 0\\ 
    \partial_t E +\nabla \cdot ((E + p)\ubf) &= 0
\end{align}
\end{subequations}
where $\rho$, $\ubf$, $p$, and $E$ are the density, velocity, pressure, and total energy density of the material. The total energy density is given by $E = \rho e+\frac{1}{2}\rho \ubf\cdot\ubf$ where $e$ is the specific internal energy density. 

We consider initial conditions relevant to ICF, that is we assume a spherical assembly of concentric shells of different material. For simplicity, we focus on a double-shell configuration, with a dense outer shell surrounding a gaseous interior. We further assume azimuthal symmetry to reduce the dynamics to two spatial dimensions (the $(r,z)$-plane) and time. We present all physical quantities in dimensionless units, implicitly in terms of reference density $\rho_{ref}$, representing the initial density of the outer material, reference length $L_{ref}$ representing the radius of the outer shell, and reference velocity $v_{ref}$ given by the speed of sound of the inner material. 

An equation of state (EOS) is used to relate $\rho$, $e$, and $p$. In the inner shell we use the SESAME tables for air \cite{osti_1487368}. In the outer material, we use the Mie-Gr\"uneisen EOS\footnote{See \cite{ward2011study} for details on differences in RMI with M-G EOS and ideal gas.}, which in dimensionless units takes the form




\begin{equation}\label{eq:MG} 
p(\rho,e) = \frac{c_s^2 \chi(\rho) (1-\frac{1}{2}\Gamma_0\chi(\rho))}{(1-s_1\chi(\rho))^2} + \Gamma_0 \rho e 
\end{equation}
with $\chi(\rho) = 1-\rho$.
The key parameters in \eqref{eq:MG} are $c_s$, the ratio of the speed of sound of the outer material to that of the inner material, $s_1$, the linear Hugoniot slope coefficient relating shock speed to particle speed in the outer material, and $\Gamma_0$, the Mie-Gr\"uneisen Gamma parameter representing the thermal pressure of vibrating atoms in the outer material. 

Fluid motion is initiated by applying a uniform inward velocity $v_0$ to the outer material, which generates a shock in the inner material. 
The shock rebounds off the origin and hits the material interface on the way out, seeding the RMI and causing small initial interface perturbations to grow. 
The resulting 2D dynamics are simulated on the unit square 
in the $(r,z)$ plane with $440$ points per dimension over the dimensionless time domain $t\in[0,1]$ 
using an arbitrary Lagrangrian-Eulerian (ALE) code CTH which employs artificial viscocity for shock capturing and adaptive mesh refinement \cite{mcglaun1990cth}. Hydrodynamic simulations were performed using Los Alamos National Laboratory's Rocinante cluster.

\subsection{Parametric setting and observed dynamics}\label{sec:data}
Our focus is to find the latent space representation of minimal dimension required to capture adequately the RMI evolution, as occurs in \eqref{eq:euler} after a shock passes through the interface between the inner and outer material (the material interface), under a range of parameters known to affect RMI growth rates. As such, we fix the Mie-Gr\"uneisen $\Gamma_0$ parameter in \eqref{eq:MG} to a nominal value since previous work has shown that it plays a minor role in the RMI evolution \cite{serino2025physics}. We thus model the solution space over variable material parameters $\mu^{mat}:= (c_s, s_1)$, consisting of the relative sound speed of the outer material and linear Hugoniot slope coefficient. The parameters $\mu^{mat}$ encode the leading order relationship between shock speed and post-shock material speed, hence directly impact RMI growth rates in the linear phase.

Along with variations in the EOS parameters $\mu^{mat}$, we explore the effect of different initial conditions on RMI evolution. We parametrize the initial conditions 
\begin{equation} \label{eq:cosine_coeff}
(\rho,\ubf,p)\big\vert_{t=0} = U_0(\mu^{IC})
\end{equation}
according to the parameter vector $\mu^{IC}:= [\mu^\CalF, v_{0}]$, which includes Fourier mode amplitudes of the initial material interface, denoted by $\mu^{\CalF}$, and the initial radial velocity of the outer shell $-v_0$. The initial interface is parametrized by
\begin{equation}\label{eq:profile_curves}
u(\theta;\mu^\CalF) = \frac{R_{in}}{8}\sum_{k=0}^8 \mu^\CalF_k \cos(2 k \theta), \quad (r(\theta),z(\theta)) = (u\cos(\theta),u\sin(\theta)), \quad \theta\in[0,\pi/2]
\end{equation}
with $R_{in} = 0.73$ 
and  $\mu^\CalF_0 = 8$ fixed throughout, such that each curve has a fixed mean radius of $R_{in}$. We refer to each such curve as a ``profile''. The list of profiles considered here is given in Table \ref{tab:initialcoeffs}, which includes a range of dominant modes and coupling strengths.

As in traditional data-driven ROM settings, we consider an observation map $\Obf$ that maps the full-order dynamics $U(t) = (\rho(t),\ubf(t),p(t))$ at parameters $\mu = [\mu^{mat},\mu^{IC}]$ to observables $\Ubf(t):=\Obf(U(t))$, emulating an experimental setting. Our goal is to construct a reduced-order model of the form
\[\frac{d}{dt}\Ubf (t) = \Fbf(\Ubf(t); \mu)\]
for the observables $\Ubf$ that accurately predicts the dynamics of $\Ubf$ at new initial conditions $\widehat{\Ubf}(0)$ and/or parameters $\widehat{\mu}$. We consider the realistic scenario of the observation map $\Obf:=\Obf_{bit}$ defined by
\begin{equation}\label{eq:bitmap}
    \Obf_{bit}(U(t)):= \ind{}(\rho_{in}(t))
\end{equation}
which maps the solution at time $t$ onto the space of binary images $\Obf_{bit}(U) \in \{0,1\}^{n_x\times n_x}$ with each pixel occupied by the inner material labeled one. Such bitmaps are extracted from the $n_x\times n_x$ corner flush with the $(r,z)$-axes and touching the origin, where we fix $n_x=70$ in the examples below. We note that $\Obf_{bit}(U(t))$ leaves many solution compartments unobserved, such as velocity and pressure fields, as well as density contours aside from the interface, although it is a weak solution to the advection equation $\partial_t U +\ubf \cdot \nabla U = 0$ with $\ubf$ the (unobserved) fluid velocity (see \ref{sec:advection}). We focus on $\Obf_{bit}$ primarily because it has been demonstrated in \cite{serino2024reconstructing} to be accessible from noisy radiographs.

\subsection{Overview of existing reduced RMI models}\label{sec:existingmodels}

Here we review theoretical progress on reduced-order modeling of RMI growth in connection with simulation and experiment, paying particular attention to ODE models. This serves to motivate our search for data-driven ROMs, as well as to demonstrate that our low-dimensional linear ODE representations do not follow readily from the literature. 

Modeling of single-mode, small-amplitude, planar interfaces is well-studied. In the linear growth phase, characterized in \cite{zhou2021rayleigh} by $a_k\leq 0.1\lambda_k$, where $\lambda_k$ is the wavelength of the dominant mode and $a_k$ is its amplitude, mode amplitudes initially grow independently, with the RTI and RMI exhibiting exponential and linear growth in time, respectively \cite{richtmyer1960taylor,meshkov1969instability}. This phase can thus be described using a system of linear ODEs for the mode amplitudes. 

 Many works have sought to quantify growth rates in the weakly nonlinear phase, which can roughly be associated with $0.1 \lambda_k<a_k <\lambda_k$ \cite{zhou2021rayleigh}. Often this can be cast in the form a nonlinear ODE. One such simple model for the bubble amplitude $h_b$ in both RTI and RMI is given by
\[\frac{d^2 h_b}{dt^2} = C_BA_og - C_D\frac{\rho_2}{\rho_2+\rho_1}\frac{(\dot{h}_b)^2}{h_b},\]
where $C_B,C_D$ are free parameters \cite{alon1995power,dimonte2006kl}. For RMI growth, the observation $h_b \sim t^\theta$, with value $\theta\approx 0.25$ found experimentally \cite{dimonte2006kl}, dictates that $C_B = 0$, which collapses the ODE above into a 1st-order nonlinear ODE $\dot{h}_b = Ch_b^{1-\frac{1}{\theta}}$ with $C$ determined by initial conditions and $\theta$ determined primarily by the Atwood number $A_o$. The spike velocity follows a power law with different $\theta$ \cite{alon1995power,dimonte2006kl}, leading to at minimum a two-dimensional ODE to describe the growth dynamics into the nonlinear regime. Other notable models include an explicit Pad\'e approximation by Zhang et al.\ \cite{zhang1996analytical} for the peak-to-trough distance in RMI, in which growth rates are found to be inverse-quadratic in time, and a parabolic approximation to the bubble profile by Goncharov \cite{goncharov2002analytical}. Both lead to 2nd-order nonlinear ODEs for interface growth which can under additional assumptions be integrated to produce highly nonlinear 1st-order ODEs. The solutions thereof were found to agree well with simulation \cite{zhang1997nonlinear} and experiment (see \cite{holmes1999richtmyer} for a review with experimental validation, and \cite{dimonte1999nonlinear} for further experimental studies). 

In a year 2000 study \cite{velikovich2000richtmyer}, it is noted that ``the linear and weakly non-linear regimes [of single-mode interfaces] are firmly established'', although the ideal case of a single-mode interface is never realized in practice (neither in experiment nor simulation), as the RMI eventually leads to a cascade of growth across many modes. A theoretical study of weak coupling between modes in RTI in the weakly nonlinear phase was offered by Haan in \cite{haan1991weakly}, in which a coupled system of nonlinear ODEs are derived to model individual Fourier mode amplitudes in the form of an asymptotic series across modes\footnote{\begin{equation}\label{eq:Haan}
\ddot{Z}_\kbf = \gamma(k)^2Z_\kbf + A k \sum_{\kbf_2} \left[\ddot{Z}_{\kbf_2} Z_{\kbf-\kbf_2}\left(1-\widehat{\kbf}_2\cdot \widehat{\kbf}\right)+\dot{Z}_{\kbf_2}\dot{Z}_{\kbf-\kbf_2}\left(\frac{1}{2}-\widehat{\kbf}_2\cdot\widehat{\kbf} - \frac{1}{2}\widehat{\kbf}_2\widehat{\kbf-\kbf_2}\right)\right] 
\end{equation}
Here $Z_\kbf$ are the mode amplitudes, $\kbf = (k_x,k_y)\in \Zbb^{2}$ are wavevectors on the 2D planar interface, $k = \|\kbf\|_2$, and $\widehat{\kbf} = \kbf/k$. The linear growth rate is given by $\gamma(k)^2 = gkA -Tk^3 / (\rho_2+\rho_1)$, which the author mentions corresponds to RMI when $g\to 0$, however a complete analysis replacing the acceleration $g$ with $g_0\delta(t)$ to fully incorporate the impulsive nature of RMI is not provided.}. The ODEs \eqref{eq:Haan} are implicit and 2nd-order, however the authors then make further approximations by evaluating higher-order terms at the linearized growth model. The dimensionality is at least the number of active modes, which is greater than the dimensionality observed in the present work. Another study derives ODEs for the coefficients $\eta_{n,n-2\ell}$ of a perturbative expansion of the interface profile $\eta(\theta,t) = \sum_{n=1}^\infty\vep^n\sum_{\ell=0}^{\lfloor n/r\rfloor} \eta_{n,n-2\ell}(t)\cos((n-2\ell)k\theta)$ as a function of angle $\theta$ in 2D convergent RMI with dominant mode $k$ \cite{wang2015weakly}. There it is argued that a 3rd-order expansion is sufficient to capture generation of second harmonics, which under additional assumptions reduces to a five-dimensional model for the five significant coefficients. The model is clearly limited to single-valued curves (functions of $\theta$), whereas our ROMs can be used to model the onset of multi-valued interfaces.


To the best of the authors' knowledge, low-dimensional ODE representations have not been derived for the RMI under strong coupling between modes and/or strongly nonlinear amplitude growth, where strongly nonlinear growth is characterized by multivalued interface curves \cite{zhou2021rayleigh}. In general, exploration of these regimes has been restricted to scientific computing \cite{zhou2020mode} and laboratory settings \cite{luo2020richtmyer}. In particular, the latter work offers a correction to the Haan model \cite{haan1991weakly} that captures a two-mode instability into the weakly nonlinear regime. This leads to a four-dimensional 2nd-order ODE (effectively eight dimensional) to capture the two dominant modes as well as their second frequency generations. Lastly, the role of the equation of state in parametrizing RMI dynamics has seen little quantitative study \cite{ward2011study,napieralski2024richtmyer}. 

The current work demonstrates that reduced-order models of low dimension should exist in the regime of multi-mode RMI in 2D convergent geometries into the nonlinear growth phase, and moreover that such ROMs can be effectively parametrized over EOS, initial interface mode amplitudes, and initial velocity. 


\section{Methodology}\label{sec:methodology}


Below we detail the main components of our parametrized ROM representation of the RMI, which is based on the LaSDI framework \cite{fries2022lasdi}. The overall objective of the ROM is to produce an approximate trajectory $\widehat{\Ubf}(t;\widehat{\mu})$ for $t\in[0,T]$ from initial conditions $\widehat{\Ubf}(0)$ and/or system parameters $\widehat{\mu}$ that matches the full-order observable $\Ubf(t,\widehat{\mu})$ given by the bitmap function \eqref{eq:bitmap}. The three main components of the ROM are the compression/decompression operators $(\phi,\psi)$, a parametrized dynamical system $\dot{z} = \fbf(z;\wbf)$ in the latent space coordinates $z \in \Rbb^{n_z}$ with coefficients $\wbf \in \Rbb^J$, and a mapping $\CalM : \Rbb^p \to \Rbb^J$ from parameters $\mu\in \Rbb^p$ to coefficients $\wbf$. These are described in Sections \ref{sec:AE}-\ref{sec:P2C}. Here, the maps $(\phi,\psi,\CalM)$ are represented by neural networks, and the requisite parameters are trained according to optimization routines described in Section \ref{sec:opt}, which include nontrivial alterations to the original LaSDI framework in order to efficiently model the evolution of a discontinuous material indicator using a high-dimensional parameter vector. 

\subsection{Compression/Decompression}\label{sec:AE}
The compression and decompression operators $\phi:\Rbb^N\to \Rbb^{n_z}$ and $\psi:\Rbb^{n_z}\to \Rbb^N$ map the ambient space $\Rbb^N$ to and from the latent space, whose dimension $n_z$ is highly problem dependent. The ultimate goal is to achieve $\psi \circ \phi \approx \text{Id}_N$. The mappings $(\phi,\psi)$ can in general be linear or nonlinear, with linear compression typically consisting of Galerkin projection onto a subspace of dominant POD modes. When the data cannot be linearly compressed to a low dimension, as is the case here
, it may still be highly compressible through nonlinear maps, which is often achieved using an autoencoder neural network architecture. 

We use autoencoder compression/decompression exclusively in this work, with $\phi = \phi(\cdot;\theta_E)$ and $\psi=\psi(\cdot;\theta_D)$ represented by fully connected feed-forward neural networks, parametrized respectively by finite-dimensional vectors $\theta_E$ and $\theta_D$. The layer widths of $\phi$ are given by $[n_x^2,n_\ell,\cdots,n_\ell,n_z]$, and reversed for $\psi$, where $n_x^2=4900 = 70\times 70$ is the number of pixels in the input image. We find that $m=2$ hidden layers with $n_\ell=500$ nodes is sufficient to capture the dynamics over the chosen scope of parameters $\mu$, where $n_\ell$ corresponds roughly to the number of POD modes needed to retain $80\%$ of the energy of the training data in the Frobenius norm\footnote{For a scalable implementation, we perform a hierarchical SVD by first extracting POD modes from each training batch of simulations that retain $90\%$ of the energy, then concatenating these batch modes into a single matrix and taking its SVD. We retain the top $n_\ell$ left singular vectors as the POD modes.}. We use the tanh activation $\sigma(x) =  \frac{e^x - e^{-x}}{e^x + e^{-x}}$ for all layers except the last layer of $\psi$, for which we use $\sigma(x) = \max(\min(x,1),0)$ to map outputs to the interval $[0,1]$, although training proceeds initially without this final activation. The outputs are thus valued in the interval $[0,1]$, which we prefer over binary-valued outputs to capture uncertainty in the material interface location.

\subsection{Latent space dynamics}\label{sec:latent}
Given an observed data snapshot $\Ubf(t)$, we define its corresponding variable in the latent space by $\Zbf(t) = \phi(\Ubf(t))$, the output of the compression operator $\phi$. The goal is to design $(\phi,\psi)$ such that $\Zbf$ approximately satisfies a parametrized ordinary differential equation 
\begin{equation}\label{eq:LSD}
    \dot{z}(t;\wbf) = \fbf(z,t; \wbf) ,\quad z(0) = z_0\in \Rbb^{n_z}, \qquad\wbf\in\Rbb^J
\end{equation}
We utilize the same methodology in the original LaSDI work, which approximates $\fbf$ from a chosen basis of vector fields, such that $\fbf$ depends linearly on the coefficients $\wbf$. Despite the multitude of nonlinear ODE representations for simplified features of RMI evolution, as reviewed in Section \ref{sec:existingmodels}, we find that linear latent space dynamics are sufficient to evolve the latent space representations. Hence, we impose linear autonomous dynamics for each $\Zbf^{[i]}$,
\[\fbf(z(t),t;\wbf^{[i]}) = \Abf^{[i]} z(t), \qquad  \wbf^{[i]} = \text{vec}(\Abf^{[i]})\] 
We expect results to improve with nonlinear latent space representations, which may be useful in  propagating the dynamics further in time. For the RMI growth regime considered here, which includes the onset of strongly nonlinear growth, it is surprising that a few-dimensional linear dynamical system is able to capture the time evolution of the material interface parametrized over material EOS parameters, initial velocity, and initial interface perturbations.  

\subsection{Parameters-to-coefficients map}\label{sec:P2C}
To test the model on a new set of parameters $\widehat{\mu}$, corresponding coefficients $\widehat{\wbf}$ must be identified to define the latent space dynamics \eqref{eq:LSD}. This is accomplished via a suitable {\it parameters-to-coefficients} map $\CalM(\mu) = \wbf$, which we represent here as an additional feedforward neural network. Previous work has included modeling $\CalM$ using Gaussian process \cite{bonneville2024gplasdi}, radial basis \cite{fries2022lasdi}, bivariate spline \cite{fries2022lasdi}, or piece-wise constant interpolation (in which regions of parameter space surrounding each training parameter $\mu^{[i]}$ are associated with the coefficient vector $\wbf^{[i]}$ \cite{fries2022lasdi}). These approaches were not found to offer successful interpolation in the current setting where a nine-dimensional parameter space is considered. (In the aforementioned works, these approaches were each presented on parameter spaces of dimension two or three.) {\it A posteriori} examination of the distribution of coefficients $\wbf$ further indicates that traditional sampling methods are inapplicable (see in Figure \ref{fig:coeff_distrib}). 

Having obtained $(\phi,\psi,\Wbf :=\{\wbf^{[i]}\}_{i=1}^S)$, where $S$ is the number of training simulations, we train $\CalM$ in a second phase. In addition to mapping parameters to coefficients $\wbf$, we also train $\CalM$ to output latent space initial conditions, that is we define
\[\CalM(\mu) = (\wbf(\mu),z_0(\mu))\]
Previous LaSDI-type applications assume that high-order initial data $\Ubf(0;\mu)$ is available, or a known parametrization thereof, from which initial conditions in the latent space $z_0$ are given simply by $\phi(\Ubf(0;\mu))$. In our setting, the RMI occurs much later in time with respect to the initial acceleration of the outer shell, hence we do not assume that the initial conditions of the data in form \eqref{eq:bitmap} are readily available. 
 We use the softplus activation for $\CalM$ and layer widths $[P,16n_z,16n_z,16n_z,16n_z,n_z^2+n_z]$, where $P$ is the dimensionality of the parameter vector $\mu$. The output dimension $n_z^2+n_z$ includes both the linear propagation matrix $\Abf(\mu)$ and latent space initial conditions $z_0(\mu)$. 

\subsection{Optimization}\label{sec:opt}

Optimization of model parameters occurs in two training phases. The first phase consists of finding optimal autoencoder parameters $(\theta_E,\theta_D)$ as well as latent space coefficients $\Wbf$ for each timeseries in the training set. The optimization problem is given by
\begin{equation}\label{eq:AEloss}
\min_{\Wbf,\theta_E,\theta_D} \CalL_{AE}(\theta_E,\theta_D;\Ubf) + \lambda_{z} \CalL_z(\Wbf;\theta_E,\Ubf) + \CalR(\Wbf,\theta_D,\theta_E)
\end{equation}
with the autoencoder loss and dynamics loss chosen here to be
\begin{equation}\label{eq:lossterms}
\CalL_{AE}(\theta_E,\theta_D) = \text{MSE}\left(\Ubf, \psi(\Zbf)\right), \quad \CalL_{z}(\Wbf) = \text{MSE}\left(\Dbf\Zbf, \Theta(\Zbf)\Wbf\right)
\end{equation}
with $\Zbf = \phi(\Ubf)$ and $\Dbf$ a suitable derivative approximation, where centered 2nd-order finite difference is used throughout. The dynamics loss couples $\Wbf$ and $\theta_E$, while the autoencoder loss $\CalL_{AE}$ couples $\theta_E,\theta_D$. 

We constrain the requisite mappings using the regularization loss $\CalR = \lambda_{\CalR_1} \CalR_1+\lambda_{\CalR_2} \CalR_2+\lambda_{\CalR_3} \CalR_3$ where each term is defined by
\begin{equation}\label{eq:reg_losses}
\CalR_1(\Wbf) = \|\Wbf\|_1, \quad \CalR_2(\theta_E,\theta_D) = \text{MSE}(\Zbf, \phi\circ \psi(\Zbf)), \quad \CalR_3(\theta_E,\theta_D) = \text{MSE}(W_E^{[0]}W_D^{[-1]}, \Ibf_{n_\ell})
\end{equation}
 Respectively, $\CalR_1$, $\CalR_2$, and $\CalR_3$ are soft constraints on the size of latent space coefficients, enforcement that the autoencoder has the projection property $P^2=P$, and that the first and last weight matrices remain an isometry between $\Rbb^N$ and $\Rbb^{n_\ell}$. Here $W_E^{[0]}$ and $W_D^{[-1]}$ are the first and last weight matrices of the encoder and decoder, respectively. 
 
 Penalization of $\CalR_1$ is useful for concentrating latent space coefficients in an $\ell_1$ ball in $\Rbb^J$, which is crucial for effectively training $\CalM$. Minimizing $\CalR_2$ implies that the flow map in the latent space and the autoencoder commute, enabling one to map to and from the latent space and remain on the same trajectory. The initialization at POD modes of the training data implies that initially $\CalR_3=0$, and maintaining small $\CalR_3$ serves to limit the extent to which outliers diverge from the true dynamics, as shown in \ref{app:PODvsrandom}. Ultimately, combining all regularization terms suffices to increase generalization over the parameter space considered, although it should be noted that a reduced-order model that is effective at modeling just the training set can be found without these terms.

The dynamics loss $\CalL_z$ is an {\it equation error}, which allows the user to avoid forward solves in the learning process and hence accelerates training. Since $\CalL_z$ is quadratic in $\Wbf$, at each optimization step the coefficients $\Wbf$ are computed via ordinary least-squares using numerical derivatives $\Dbf\Zbf$ and basis functions $\Theta(\Zbf)$, and back-propagation is performed through the pseudo-inverse $\Theta(\Zbf)^\dagger$, which depends implicitly on $\theta_E$. This is in contrast to {\it output error} base approaches, such as Neural ODEs (NODE), which require solving forward in time the candidate dynamical system at each iteration \cite{chen2018neural,lee2021parameterized}. We propose to incorporate an output error loss only in the next phase, where far fewer neural network weights are optimized. 


The second phase consists of training the parameters-to-coefficients map $\CalM$, whose free parameters we denote by $\theta_\mu$. Here the optimization problem is 
\begin{equation}\label{eq:P2Closs}
\min_{\theta_\mu} \text{MSE}((\Wbf,\Zbf_0), \CalM(\pmb{\mu};\theta_\mu)) + \lambda_{\Ubf_0} \text{MSE} (\Ubf_0, \psi(\Zbf_0(\pmb{\mu}))) + \lambda_{Z(T)} \text{MSE}(\Zbf(T),\exp(\Abf T)\Zbf_0(\pmb{\mu}))
\end{equation}
where $\pmb{\mu}$ denotes the collection of parameters corresponding to training set simulations, $\Zbf_0(\mu)$ denotes the initial latent space variables output by $\CalM$, and $\Ubf_0$ denotes the first observed variables of each training simulation. The loss terms here enforce agreement with the least-squares (equation-error) optimal coefficients $\Wbf$ and the latent space initial conditions $\Zbf_0$, agreement with the corresponding ambient space initial conditions $\Ubf_0$, and a single output error loss penalizing the difference between the final time in the latent space $\Zbf(T)$ and the model prediction $\exp(\Abf T)\Zbf_0(\pmb{\mu})$. While previous LaSDI applications focused only on the equation error loss, in our setting of low-dimensional linear dynamics, a single evaluation of the matrix exponential is inexpensive, especially when appending to the parameters-to-coefficients loss rather than the autoencoder loss, since $\CalM$ has far fewer parameters than $(\phi,\psi)$. We also find that the second loss terms improves accuracy due to the close proximity of initial conditions across simulations.

For training, we use the first-order optimization method Adam throughout, with a batch size of 32 simulations and an initial learning rate of $\ell_0 = 0.1$. We implement a learning rate schedule as well as early stopping for both the autoencoder and the parameters-to-coefficients map. In training the former, after 500 epochs we decrease the learning rate according to $\ell_{n+1} = 0.995\ell_n$ every $10$th epoch. When the total loss falls below $0.01$, we append to the autoencoder the activation function $\sigma(x) = \min\{1,\max\{0,x\}\}$ after the final layer, mapping pixels in the output of the decoder $[0,1]$. We find that training is greatly accelerated by initially excluding this outer activation function. We choose loss weights $\lambda_z = 0.2$, noting that the dynamics loss $\CalL_z$ equilibrates much faster than the autoencoder loss $\CalL_{AE}$, and regularization weights $\lambda_{R_1} = 1$e-$4$, $\lambda_{R_1} = 0.1$, $\lambda_{R_1} = 0.1$.
In training $\CalM$, we use $\lambda_{\Ubf_0}=\lambda_{Z(T)}=0.1$. 

For early stopping, we implement a warning-weight-halt strategy, consisting of a binary indicator $I_{warn}$ based on model performance on a hold-out validation dataset. When $I_{warn}$ returns \texttt{true}, a weighting period is initialized of duration $\tau_\text{wait}$ epochs. If $I_{warn}$ still returns true $\tau_\text{wait}$ epochs later, optimization is halted. For the autoencoder we use 
\[I^{AE}_{warn}(n) = \left\{\CalL^n_{AE}(\Ubf^{val}) > \tau_{warn}\min_n\CalL^n_{AE}(\cdot; \Ubf^{val})\right\}\ \text{or}\ \left\{\CalL^n_{z}(\Ubf^{val}) > \tau_{warn}\min_n\CalL^n_{z}(\cdot; \Ubf^{val})\right\} \]
where $\Ubf^{val}$ denotes the validation dataset and $\CalL^n$ denotes the corresponding loss term at the $n$th epoch. Similarly, we have the warning function for the parameters-to-coefficients map 
\[I^{P2C}_{warn} = \left\{\text{MSE}((\Wbf^{val},\Zbf_0^{val}),\CalM_n(\pmb{\mu}^{val})) > \tau_{warn}\min_n\text{MSE}((\Wbf^{val},\Zbf^{val}_0),\CalM_n(\pmb{\mu}^{val}))\right\}\]
In both cases we let $\tau_{warn} = 1.25$ and $\tau_{wait} = 200$ epochs.

\subsubsection{Latent space dimension}
Of primary interest in the current study is the dimensionality of the latent space, $n_z$. In initial studies (not shown) we trained LaSDI models to reproduce training datasets using $n_z=10$. This led to over-parametrization as seen by the autoencoder weights, with many entries unchanged during training. Using the Haan multimodal model \eqref{eq:Haan} for guidance, we then restricted to $n_z = 6$, the number of active Fourier modes at the interface. From this it was observed that multiple latent space trajectories were constant in time, indicating a redundant latent representation. By exploring LaSDI models with $n_z<6$, we found $n_z=3$ to be the minimum dimension able to both reproduce the data to sufficient accuracy and generalize over the test set. For brevity we include only results for $n_z\in\{2,3,4\}$ below, under different restrictions in parameter space, finding that $n_z=3$ strikes the right balance between efficiency and generalization. In addition, \ref{sec:latentspace_vis} includes visualizations of the latent space for $n_z=3$, indicating that simulations cluster according to their dominant interface mode (Fig. \ref{fig:latentspace}), an observation we aim to incorporate into much future studies with larger interface parameter spaces. 

\subsubsection{POD initialization}
While linear compression is insufficient for achieving low-dimensional RMI representations, we include in \ref{app:PODvsrandom} evidence that initializing the first hidden layer weights of the encoder $\phi$ and last hidden layer weights of the decoder $\psi$ at a fixed number of POD modes of the data offers substantial improvements in training. We compare this POD initialization to random initialization in Figure \ref{fig:randinit}, finding that the former reduces the number of outliers when applying the ROM to out-of-sample data and reduces training time. In effect, POD initialization with the Adam optimizer allows AE weights to follow the steepest descent (up to minibatching) to a nonlinear manifold from the best linear manifold representation of the data.


\section{Results}\label{sec:results}

\begin{figure}
    \includegraphics[trim={30 320 70 0},clip,width=1\textwidth]{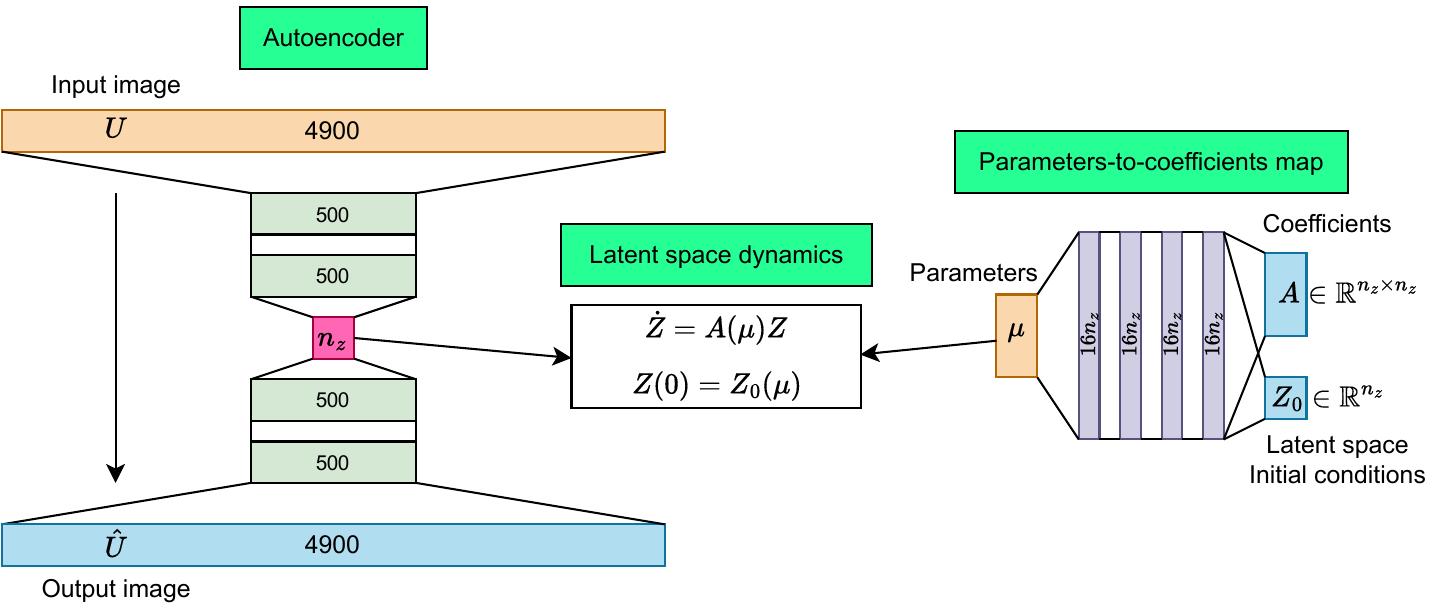} \\
\caption{Diagram of parametrized reduced-order model, consisting of a shallow autoencoder network with two hidden layers in the encoder $\phi$ and decoder $\psi$, an expansive parameters-to-coefficients network $\CalM$ with four hidden layers, and a parametrized linear latent-space dynamical system. The layer widths are indicated in each layer, and $n_z \in \{2,3,4\}$ is examined.}
\label{fig:networkdiagram}
\end{figure}

\begin{table}
\begin{center}
\begin{tabular}{|l|c|*{5}{c|}}
\hline
$v_{0}$ & $2.845$ & $2.865$ & $2.880$ & $2.890$ \\
$s_1$ & $1.220$ & $1.342$ & $1.464$ & \\
$c_s$ & $9.204$ & $10.23$ &  $10.71$ & $11.25$ \\
\hline
\end{tabular}
\end{center}
\caption{Non-interface parameters. The initial velocity $v_{0}$ and outer material sound speed $c_s$ are relative to the sound speed of the inner material.}
\label{tab:matparams}
\end{table}

\begin{table}
\begin{center}
\begin{tabular}{|l|c|*{8}{c|}}
\hline
Profile & $\mu^\CalF_0$ & $\mu^\CalF_1$ & $\mu^\CalF_2$ & $\mu^\CalF_3$ & $\mu^\CalF_4$ & $\mu^\CalF_5$ & $\mu^\CalF_6$ & $\mu^\CalF_7$ & $\mu^\CalF_8$ \\
\hline
1 & 8 & 0 & 0 & 0 & 0 & 0 & 0 & 0 & 0.08 \\
2 & 8 & 0 & 0 & 0 & 0.08 & 0 & 0 & 0 & 0 \\
3 & 8 & 0 & 0.08 & 0 & 0 & 0 & 0 & 0 & 0 \\
4 & 8 & 0 & 0 & 0 & 0 & 0 & 0 & 0 & 0.075 \\
5 & 8 & 0 & 0 & 0 & 0.075 & 0 & 0 & 0 & 0 \\
6 & 8 & 0 & 0.075 & 0 & 0 & 0 & 0 & 0 & 0 \\
7 & 8 & 0 & 0.0075 & 0 & 0 & 0 & 0.0025 & 0 & 0.065 \\
8 & 8 & 0.0075 & 0 & 0.0025 & 0.065 & 0 & 0 & 0 & 0 \\
9 & 8 & 0.005 & 0.0657 & 0 & 0 & 0 & 0 & 0 & 0 \\
10 & 8 & 0 & 0 & 0 & 0 & 0 & 0 & 0 & 0.06 \\
11 & 8 & 0 & 0 & 0 & 0.06 & 0 & 0 & 0 & 0 \\
12 & 8 & 0 & 0.06 & 0 & 0 & 0 & 0 & 0 & 0 \\
13 & 8 & 0 & 0 & 0 & 0 & 0 & 0 & 0 & 0.055 \\
14 & 8 & 0 & 0 & 0 & 0.055 & 0 & 0 & 0 & 0 \\
15 & 8 & 0 & 0.055 & 0 & 0 & 0 & 0 & 0 & 0 \\
16 & 8 & 0 & 0.0075 & 0 & 0 & 0 & 0.0025 & 0 & 0.045 \\
17 & 8 & 0.0075 & 0 & 0.0025 & 0.045 & 0 & 0 & 0 & 0 \\
18 & 8 & 0.0051 & 0.0457 & 0 & 0 & 0 & 0 & 0 & 0 \\
19 & 8 & 0 & 0 & 0 & 0.04 & 0 & 0 & 0 & 0 \\
20 & 8 & 0 & 0.04 & 0 & 0 & 0 & 0 & 0 & 0 \\
21 & 8 & 0 & 0 & 0 & 0 & 0 & 0.08 & 0 & 0 \\
22 & 8 & 0 & 0 & 0 & 0 & 0 & 0.075 & 0 & 0 \\
23 & 8 & 0.0075 & 0 & 0.0025 & 0 & 0 & 0.065 & 0 & 0 \\
24 & 8 & 0 & 0 & 0 & 0 & 0 & 0.06 & 0 & 0 \\
25 & 8 & 0 & 0 & 0 & 0 & 0 & 0.055 & 0 & 0 \\
26 & 8 & 0.0075 & 0 & 0.0025 & 0 & 0 & 0.045 & 0 & 0 \\
27 & 8 & 0 & 0.04 & 0 & 0.04 & 0 & 0 & 0 & 0 \\
28 & 8 & 0 & 0.02 & 0 & 0.06 & 0 & 0 & 0 & 0 \\
29 & 8 & 0 & 0.06 & 0 & 0.02 & 0 & 0 & 0 & 0 \\
30 & 8 & 0.0075 & 0.04 & 0.0025& 0.04 & 0 & 0 & 0 & 0 \\
31 & 8 & 0.0075 & 0.02 & 0.0025& 0.06 & 0 & 0 & 0 & 0 \\
32 & 8 & 0.0075 & 0.06 & 0.0025& 0.02 & 0 & 0 & 0 & 0 \\
\hline
\end{tabular}
\end{center}
\caption{Cosine coefficients defining each initial material interface profile, according to \eqref{eq:cosine_coeff}.}
\label{tab:initialcoeffs}
\end{table}

Below we demonstrate the performance of the proposed reduced-order modeling framework in simulating the dynamics of a material interface undergoing RMI for various coupling strengths and extents in parameter space. From this we conclude that the dynamics are most effectively modeled in a three-dimensional latent space ($n_z=3$). 
In Sections \ref{sec:weakcoupling}, \ref{sec:strongcoupling}, and \ref{sec:allcoupling}, we examine three interface regimes: (1) weak coupling only, such that a dominant wavemode can be associated with each simulation, (2) strong coupling between two modes, and (3) a dataset combining the two cases. We include visualizations of the latent space for $n_z=3$ in Figure \ref{fig:latentspace} of \ref{sec:latentspace_vis}.

Data is extracted from the last 41 snapshots of the density field ($t\in[0.9,1.0]$) over the first $70\times70$ gridpoints, capturing the material interface from just before shock passage into the nonlinear RMI growth phase. The material and initial conditions parameters are varied according to Tables \ref{tab:matparams} and  \ref{tab:initialcoeffs}, totaling 1536 simulations over a nine-dimensional parameter space. For each experiment below, we randomly split the data into training-validation-testing sets of proportion $0.725$-$0.025$-$0.25$, leading to training sets of size 905, 696, and 1114 simulations for the weak coupling, strong coupling, and mixed experiments, respectively. We restrict our focus to the architecture in Figure \ref{fig:networkdiagram}, with the first and last weight matrices of the autoencoder initialized at the top 500 POD modes as outlined above. Results are reported for the best performing model out of three training stages, each with randomly initialized network weights (apart from the POD modes) and randomly selected training datasets.

\subsection{Performance Metrics}

Performance is measured using two relative metrics, the mean pixel error (MPE) and the Jaccard Loss (JL). For images $A$ and $B$, these are defined as follows:
\begin{subequations}\label{eq:metrics}
\begin{align}
  \text{MPE}(A,B) &:= \frac{1}{N}\sum_{i=1}^N|A_i-B_i|\label{eq:MPE} \\
  \text{JL}(A,B) &:= 1-\frac{\#\{A\cap B\}}{\#\{A\cup B\}} \label{eq:JL}
\end{align}
\end{subequations}
Here $N=4900$ is the total number of pixels, $\#\{S\}$ counts the number of 1's in a binary image $S$.
For JL, images are converted to binary by assigning 1's to all nonzero values. The MPE makes no such assumption, using the ROM output onto values in the interval $[0,1]$. 
The Jaccard loss (1-Jaccard Score) is a relative measure of similarity, with JL$(A,B) \in [0,1]$ and $ \text{JL}(A,B) = 0$ when $A=B$. For binary images, $100\times  \text{JL}(A,B)\%$ is the percent of mismatched pixels between $A$ and $B$ out of the total number of 1's across $A$ and $B$. 

For each metric we report the minimum, median, and maximum values over the test set as a function of time. We also report and visualize the 95th percentile of the maximum-in-time error over the observed 41 snapshots, denote this by $\texttt{err}$, such that the model performs better than $\texttt{err}$ on 95\% of test cases. 

\subsection{AE-ROM, V-ROM, F-ROM}

The proposed ROM framework incorporates three levels of modeling that may be useful in different contexts. Each level reveals different aspects of the intrinsic dimensionality of the RMI solution space and its ability to be faithfully parametrized using parameters in Tables \ref{tab:matparams} and Table \ref{tab:initialcoeffs}. The first level is the {\it static autoencoder} model, referred to as the AE-ROM, which consists solely of the trained autoencoder. The utility of the AE model is in compressing the data and revealing the intrinsic dimensionality of the set of snapshots. Successful performance of the AE model on a test set indicates that the $n_z$-dimensional manifold faithfully represents the data, but says nothing about its dynamics. This can be compared to traditional dimensionality reduction such as POD or local PCA, or with dimensionality estimation methods (for a review see \cite{lee2007nonlinear}). We note that none of the available techniques in the python package \texttt{scikit-dimension} could corroborate our finding that the RMI is effectively three dimensional \cite{bac2021scikit}, with the low end of dimensionality estimates producing $n_z\sim 12$ on limited portions of the data (dimensionality estimates of the entire dataset often exceeded the computation time required to train the ROM on the same dataset). 

The next level of modeling considered is the {\it validation model} or V-ROM, which consists of the autoencoder and latent space dynamics but excludes the parameters-to-coefficients map $\CalM$. The nomenclature ``validation'' is used because the V-ROM validates all assumptions on the solution space, namely that it can be compressed to an $n_z$-dimensional manifold over which it obeys a dynamical system with functional form given by \eqref{eq:LSD}. To use the V-ROM, one takes an available simulation $\Ubf^*$ (implicitly at parameters $\mu^*$), compresses it into the latent space using the encoder $\phi$ to arrive at a latent space trajectory $\Zbf^*$, then learns the coefficients $\wbf^*$ of the dynamical system by minimizing the equation error loss $\CalL_z$ \eqref{eq:lossterms}, a least-squares solve. The output of the V-ROM then includes simulating the resulting dynamical system and decoding it with $\psi$. The V-ROM can take in different data than it was trained on, and may only need several snapshots to estimate $\wbf^*$. We do not explore this flexibility here, using all 41 snapshots of each simulation to estimate coefficients in V-ROM models and assessing their ability to reproduce the observed snapshots. 

Lastly, the F-ROM represents the full reduced-order model we seek, which takes in parameters $\mu$ and simulates the entire dynamics over the observed time interval $t\in[0.9,1.0]$ 
using the parameters-to-coefficients map $\CalM$, the latent dynamical system, and the decoder $\psi$. We focus mainly on the performance of the F-ROM, noting that performance typically degrades from the V-ROM to the F-ROM, highlighting the challenges in learning the parametric dependence via $\CalM$.

\subsection{Experiment 1: Weak coupling}\label{sec:weakcoupling}

Our first experiment treats the weak mode-coupling regime, where we restrict the ROM to modeling profiles 1-26 in Table \ref{tab:initialcoeffs}. The scope of the ROM thus includes capturing the RMI evolution under a total of four dominant wavemodes $k\in \{2,4,6,8\}$, each of which is varied over three initial amplitudes and three coupling strengths to other modes, as well as the parameters $(v_{0},s_1,c_s)$ given in Table \ref{tab:matparams}.

\begin{figure}
\begin{tabular}{@{}l@{}}
    \includegraphics[trim={110 0 100 0},clip,width=1\textwidth]{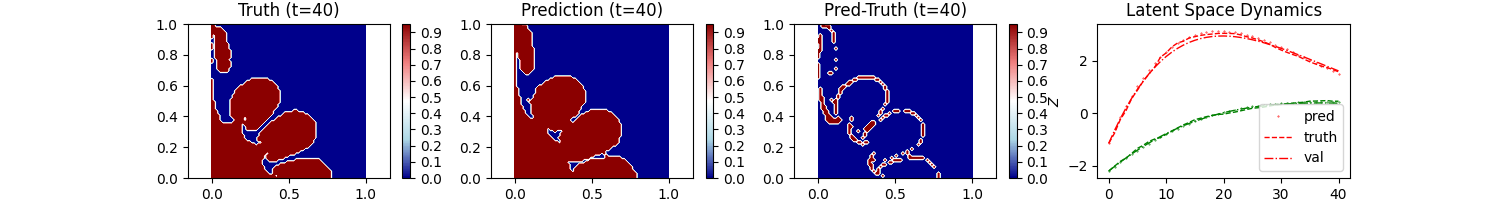} \\
    \includegraphics[trim={110 0 100 0},clip,width=1\textwidth]{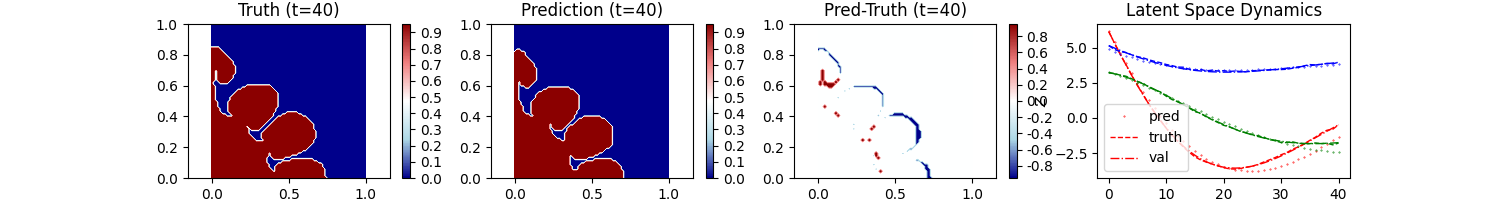} \\
    \includegraphics[trim={110 0 100 0},clip,width=1\textwidth]{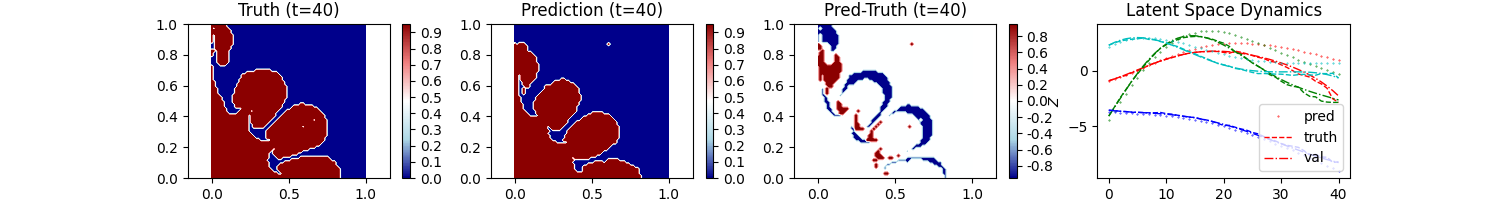}
\end{tabular}
\begin{center}
{\footnotesize

    \begin{tabular}{c|cc|cc}
        $n_z$ & MPE (AE-ROM, tr.) & JL (AE-ROM, tr.) & MPE (AE-ROM, te.) & JL (AE-ROM, te.)  \\ \hline
        2 & 0.041 & 0.10  & 0.043 & 0.12 \\
        3 & 0.019 & 0.05 & 0.026 & 0.065   \\
        4 & 0.015 & 0.039 & 0.025 & 0.061 \\ \hline
        $n_z$ & MPE (V-ROM, tr.) & JL (V-ROM, tr.) & MPE (V-ROM, te.) & JL (V-ROM, te.)  \\ \hline
        2 & 0.097 & 0.28  & 0.097 & 0.27 \\
        3 & 0.025 & 0.065 & 0.031 & 0.078   \\
        4 & 0.018 & 0.057 & 0.038 & 0.10 \\ \hline
        $n_z$ & MPE (F-ROM, tr.) & JL  (F-ROM, tr.) & MPE  (F-ROM, te.) & JL (F-ROM, te.) \\ \hline
        2 & 0.098 & 0.28 & 0.10 & 0.28 \\
        3 & 0.028 & 0.078 & 0.037 & 0.099  \\
        4 & 0.034 & 0.10  & 0.075 & 0.21  \\ \hline
    \end{tabular}
}
\end{center}
\caption{ROM performance over the weak coupling regime: profiles 1-26 over parameters $[v_{0},s_1,c_s]$ (see Tables \ref{tab:initialcoeffs}, \ref{tab:matparams}). Tables include 95th percentile mean pixel error (MPE) and Jaccard loss (JL) over the training (tr.) and testing (te.) sets for models with latent space dimension $n_z\in \{2,3,4\}$. Images depict final-time model outputs ($t=40$ indicating the final snapshot) for testing simulations at the 95th percentile in JL using (top to bottom) $n_z = \{2,3,4\}$ (i.e.\ the model performed better than the depicted case on 95\% test cases). Left to right: ground truth, ROM prediction, difference between predicted and truth, and latent space dynamics over the time interval. The latent space trajectories are plotted for the AE, V-ROM, and F-ROM models in dashed, dot-dashed, and dotted lines, respectively, with colors representing different latent space coordinates over time.}
\label{fig:weakcoupling}
\end{figure}

Figure \ref{fig:weakcoupling} includes the 95th percentile scores for the AE-ROM, V-ROM, and F-ROM over the training and test datasets. Images of the final-time model outputs are shown for the simulations at the 95th percentile within each test set, including the corresponding latent space trajectories, for the $n_z=2,3,4$ models (top to bottom). The final column indicates that the F-ROM (labeled `pred') produces latent space trajectories that deviate slightly in $n_z=3$ and $n_z=4$ from the AE-ROM and V-ROM models (labeled `truth' and `val'), which are overlapping. Here, the $n_z=2$ dimensional AE-ROM can adequately compress the data, yielding 95th percentile MPE and JL of $4.3\%$ and $12\%$, but the V-ROM and F-ROM lead to large errors. Latent space dimension $n_z=4$ provides accurate AE-ROM and V-ROM models, indicating that the dynamics can be adequately represented as a 4-dimensional linear model, but fails to generalize in the F-ROM, indicating that the parametric dependence is obscured with $n_z=4$. The $n_z=3$ model leads to a successful F-ROM, which generalizes to the test set with a 95th percentile MPE and JL of 3.7\% and 9.9\%, which leads to minor visible differences in prediction of the final-time RMI profile (see middle row of Fig. \ref{fig:weakcoupling}).   

\begin{figure}
\begin{center}
\begin{tabular}{@{}c@{}c@{}c@{}}
\fbox{$n_z=2$} & \fbox{$n_z=3$}& \fbox{$n_z=4$} \\
        \includegraphics[trim={0 20 0 0},clip,width=0.33\textwidth]{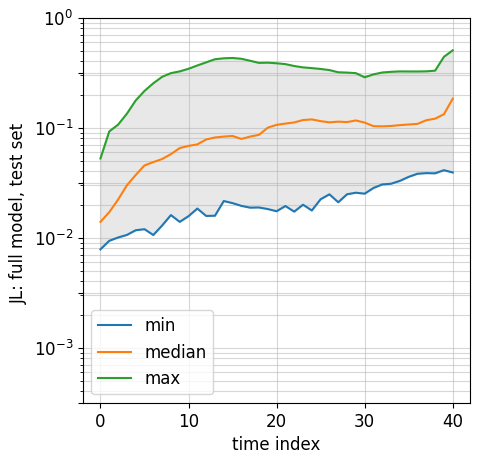} &
    \includegraphics[trim={0 20 0 0},clip,width=0.33\textwidth]{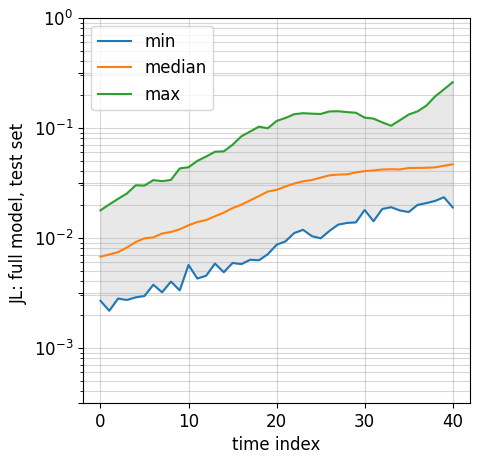} &
    \includegraphics[trim={0 20 0 0},clip,width=0.33\textwidth]{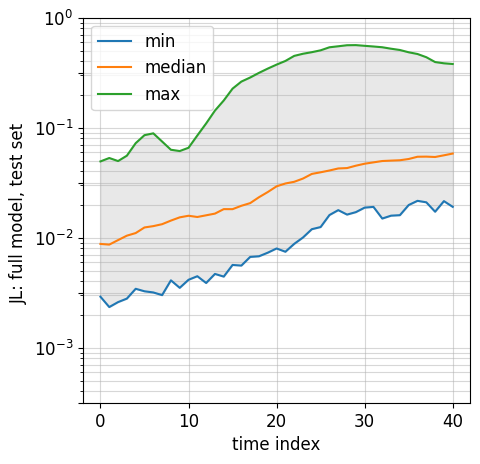} \\ \hline
        \includegraphics[trim={0 6 0 0},clip,width=0.33\textwidth]{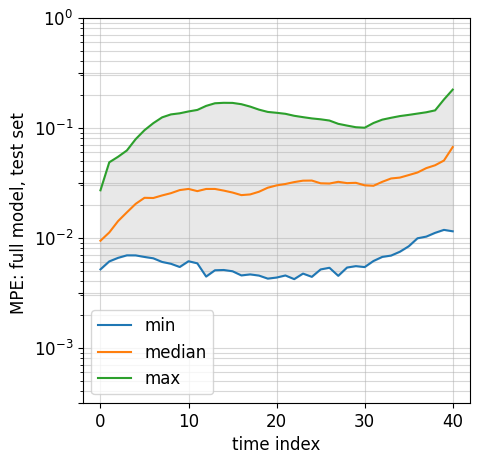} &
    \includegraphics[trim={0 6 0 0},clip,width=0.33\textwidth]{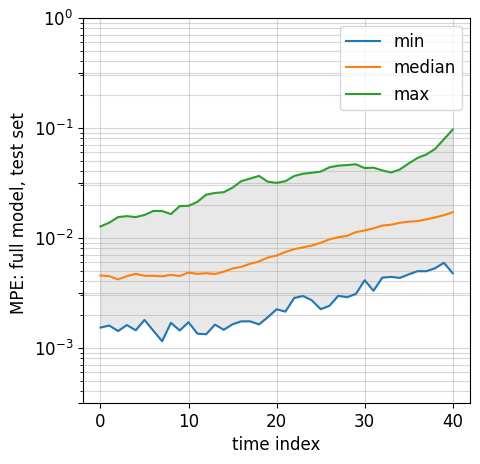} &
    \includegraphics[trim={0 6 0 0},clip,width=0.33\textwidth]{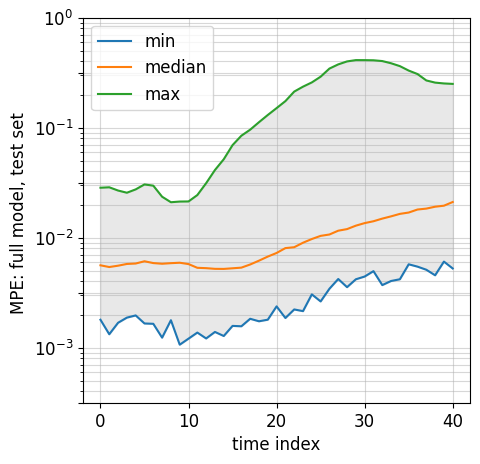} 
\end{tabular}
\end{center}
\caption{Performance of F-ROMs in modeling RMI under weak mode coupling according to the minimum, median, and maximum JL (top) and MPE (bottom) metrics \eqref{eq:metrics} over the test set as a function of time.}
\label{fig:weakcoupling_minmax}
\end{figure}

The F-ROM performance is examined in more detail in Figure \ref{fig:weakcoupling_minmax}, which plots the minimum, median, and maximum values of each metric on the test set over the course of the simulation. The $n_z=3$ model is most accurate and exhibits lowest variance, with errors clustering around the median. Median performance is similar for the $n_z=4$ model but with higher variance, indicating that increasing the dimension reduces generalizability. Errors generally increase over time for the $n_z=3$ model, with a sharp uptick in the last 6 snapshots. This can be associated with increased complexity of the interface curves at the onset of strongly nonlinear growth. Beyond this point, it may be necessary to change the model architecture, and in general we do not expect to model the strongly nonlinear growth phase using only the first 9 interface harmonics as considered here (i.e.\ Table \ref{tab:initialcoeffs}).

\subsection{Experiment 2: Strong coupling}\label{sec:strongcoupling}

We now focus on modeling the strong coupling regime between modes $k=2$ and $k=4$ by including profiles 27-32 and omitting profiles with dominant modes $k=6$ or $k=8$ (profiles [1,4,7,10,13,16,21-26]). The resulting 20 profiles capture a variety of couplings between $k=2$ and $k=4$ modes with amplitudes $(\mu^\CalF_2,\mu^\CalF_4)\in (0,x),(0.02,0.06),(0.04,0.04),(0.06,0.02),(x,0)$ for $x\in[0.04,0.08]$. 

\begin{figure}
\begin{tabular}{@{}l@{}}
    \includegraphics[trim={110 0 100 0},clip,width=1\textwidth]{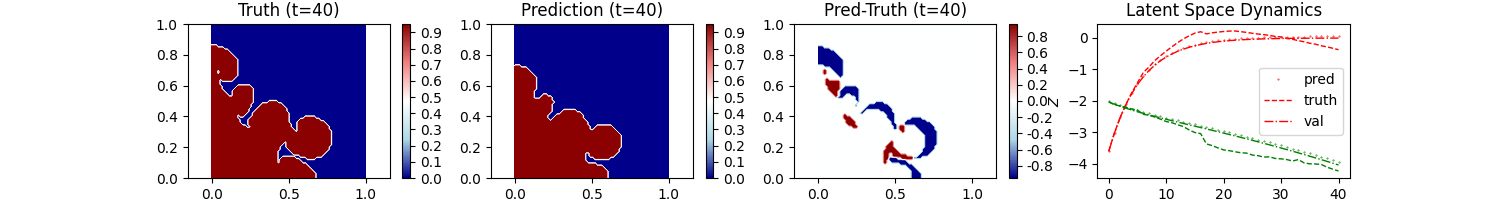} \\
    \includegraphics[trim={110 0 100 0},clip,width=1\textwidth]{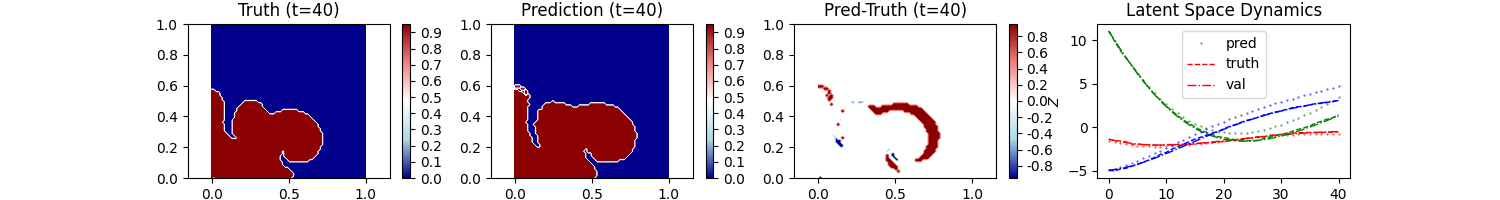} \\
    \includegraphics[trim={110 0 100 0},clip,width=1\textwidth]{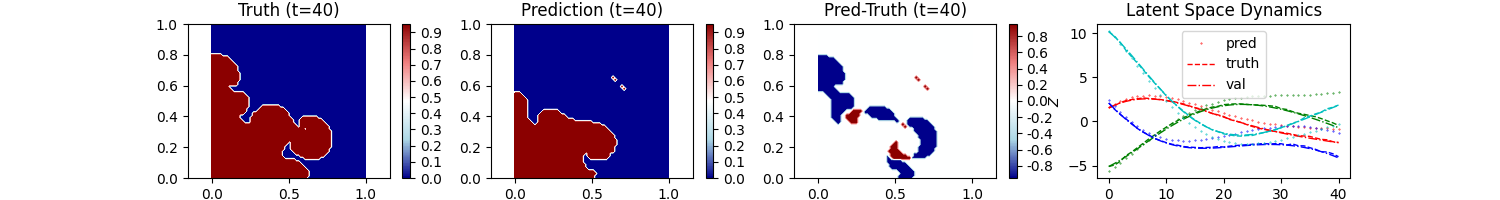}
\end{tabular}
\begin{center}
{\footnotesize

    \begin{tabular}{c|cc|cc}
        $n_z$ & MPE (AE-ROM, tr.) & JL (AE-ROM, tr.) & MPE (AE-ROM, te.) & JL (AE-ROM, te.)  \\ \hline
        2 & 0.027 & 0.071 &  0.037 & 0.098   \\
        3 & 0.018 & 0.048 & 0.025 & 0.067   \\
        4 & 0.017 & 0.042 & 0.027 & 0.067 \\ \hline
        $n_z$ & MPE (V-ROM, tr.) & JL (V-ROM, tr.) & MPE (V-ROM, te.) & JL (V-ROM, te.)  \\ \hline
        2 & 0.070 & 0.22 &  0.072 & 0.22   \\
        3 & 0.020 & 0.060 & 0.027 & 0.076   \\
        4 & 0.017 & 0.045 & 0.031 & 0.082 \\ \hline
        $n_z$ & MPE (F-ROM, tr.) & JL  (F-ROM, tr.) & MPE  (F-ROM, te.) & JL (F-ROM, te.) \\ \hline
        2 & 0.069 & 0.22  & 0.071 & 0.22 \\
        3 & 0.022 & 0.068  & 0.033 & 0.088 \\
        4 & 0.028 & 0.084  & 0.051 & 0.13 \\ \hline
    \end{tabular}
}
\end{center}
\caption{95th percentile scores for modeling strong coupling between $k=2$ and $k=4$ modes, with variable $v_{0}$ and EOS parameters $[s_1,c_s]$. Figures are analogous to those in \ref{fig:weakcoupling}.}
\label{fig:strongcoupling}
\end{figure}

\begin{figure}
\begin{center}
\begin{tabular}{@{}c@{}c@{}c@{}}
\fbox{$n_z=2$} & \fbox{$n_z=3$}& \fbox{$n_z=4$} \\
        \includegraphics[trim={0 20 0 0},clip,width=0.33\textwidth]{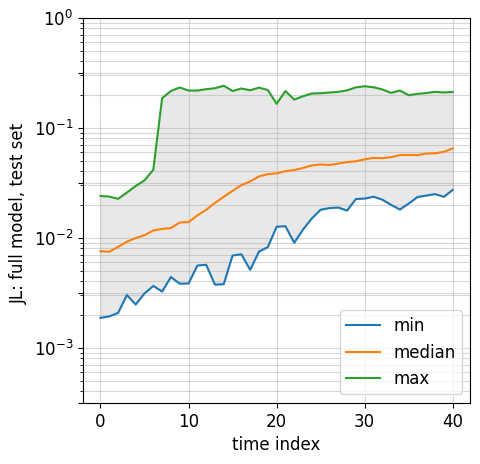} &
    \includegraphics[trim={0 20 0 0},clip,width=0.33\textwidth]{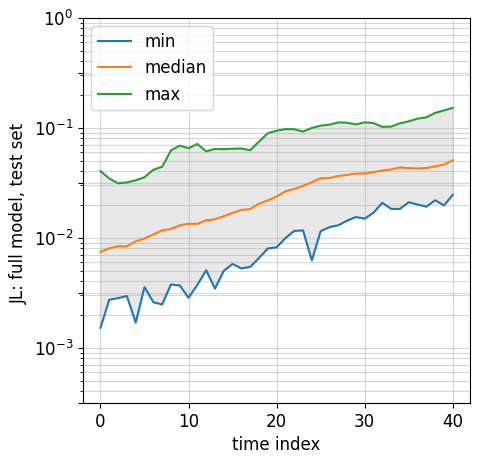} &
    \includegraphics[trim={0 20 0 0},clip,width=0.33\textwidth]{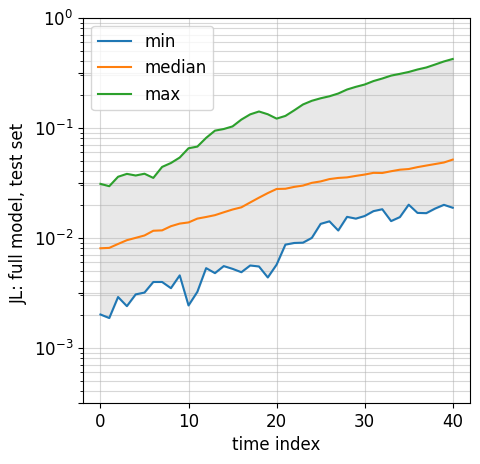} \\ \hline
        \includegraphics[trim={0 6 0 0},clip,width=0.33\textwidth]{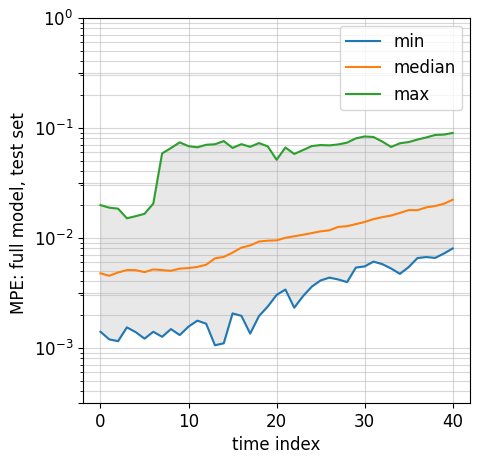} &
    \includegraphics[trim={0 6 0 0},clip,width=0.33\textwidth]{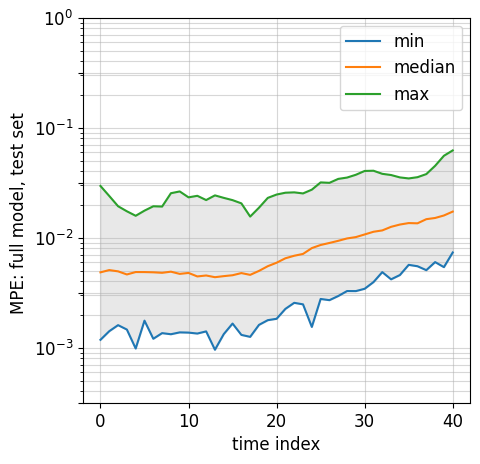} &
    \includegraphics[trim={0 6 0 0},clip,width=0.33\textwidth]{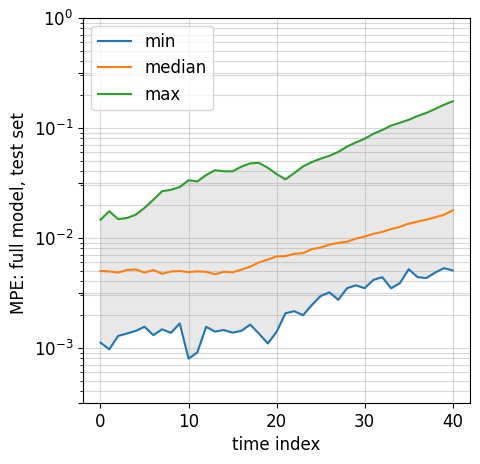} 
\end{tabular}
\caption{Performance of F-ROMs in modeling RMI under strong mode coupling according to the minimum, median, and maximum JL (top) and MPE (bottom) metrics \eqref{eq:metrics} over the test set as a function of time.}
\label{fig:strongcoupling_minmax}
\end{center}
\end{figure}

While theoretical results would predict an increase in dimensionality, the trend is similar to the previous experiment, although the $n_z=2$ latent space trajectories in Figure \ref{fig:strongcoupling} indicate that the two-dimensional embedding can no longer adequately be modeled by a linear dynamical system, as expected. The $n_z=3$ F-ROM model again outperforms that of $n_z=4$, although the performance of the corresponding V-ROMs are nearly identical. The min-median-max F-ROM test errors in Figure \ref{fig:strongcoupling_minmax} again indicate that $n_z=4$ leads to increased variance in performance and more outliers later in time.

\subsection{Experiment 3: All profiles}\label{sec:allcoupling}

\begin{figure}
\begin{tabular}{@{}l@{}}
    \includegraphics[trim={110 0 100 0},clip,width=1\textwidth]{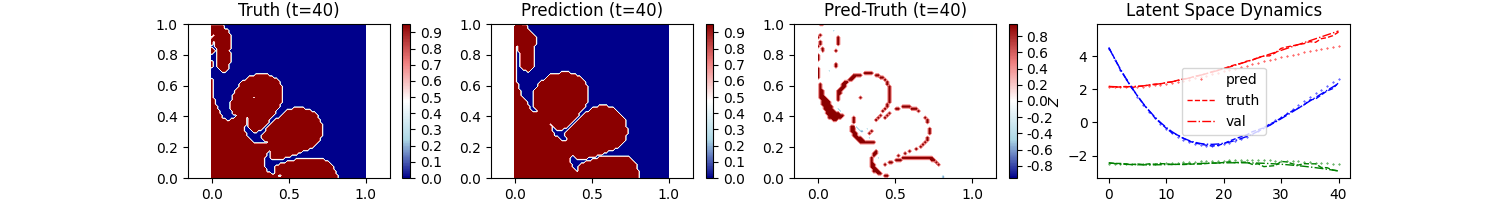} \\
    \includegraphics[trim={110 0 100 0},clip,width=1\textwidth]{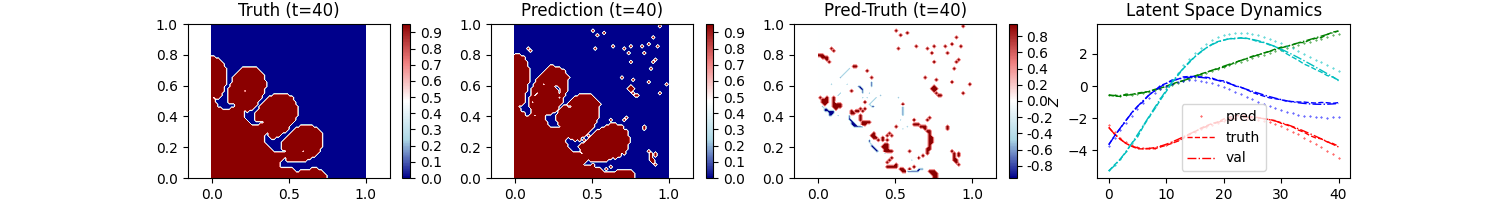}
\end{tabular}
\begin{center}
{\footnotesize

    \begin{tabular}{c|cc|cc}
        $n_z$ & MPE (AE-ROM, tr.) & JL (AE-ROM, tr.) & MPE (AE-ROM, te.) & JL (AE-ROM, te.)  \\ \hline
        3 & 0.022 & 0.056 & 0.028 & 0.072   \\
        4 & 0.016 & 0.043 & 0.024 & 0.062 \\ \hline
        $n_z$ & MPE (V-ROM, tr.) & JL (V-ROM, tr.) & MPE (V-ROM, te.) & JL (V-ROM, te.)  \\ \hline
        3 & 0.027 & 0.075 & 0.032 & 0.091   \\
        4 & 0.021 & 0.057 & 0.026 & 0.069 \\ \hline
        $n_z$ & MPE (F-ROM, tr.) & JL  (F-ROM, tr.) & MPE  (F-ROM, te.) & JL (F-ROM, te.) \\ \hline
        3 & 0.028 & 0.09  & 0.035 & 0.10  \\
        4 & 0.035 & 0.096  & 0.056 & 0.14 \\ \hline
    \end{tabular}
}
\end{center}
\caption{95th percentile scores for modeling all profiles in Table \ref{tab:initialcoeffs}. Figures are analogous to those in \ref{fig:weakcoupling}, with the $n_z=2$ omitted.}
\label{fig:allcoupling}
\end{figure}

\begin{figure}
\begin{center}
\begin{tabular}{@{}c@{}c@{}}
\fbox{$n_z=3$}& \fbox{$n_z=4$} \\
    \includegraphics[trim={0 20 0 0},clip,width=0.33\textwidth]{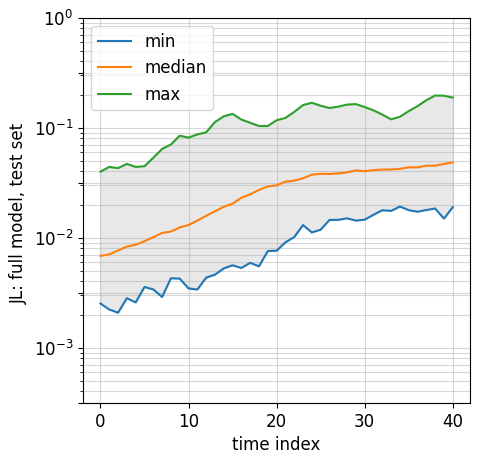} &
    \includegraphics[trim={0 20 0 0},clip,width=0.33\textwidth]{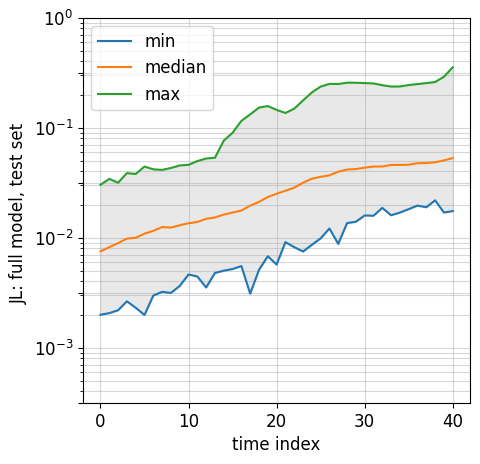} \\ \hline
    \includegraphics[trim={0 6 0 0},clip,width=0.33\textwidth]{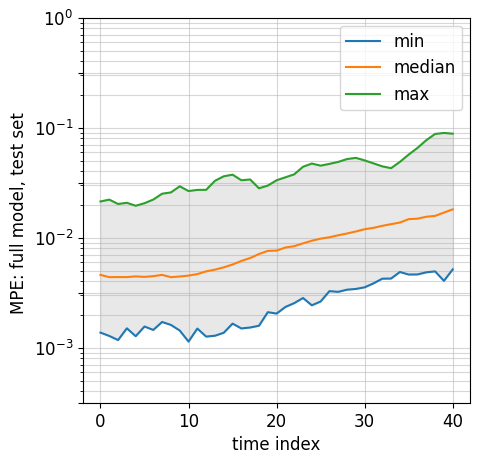} &
    \includegraphics[trim={0 6 0 0},clip,width=0.33\textwidth]{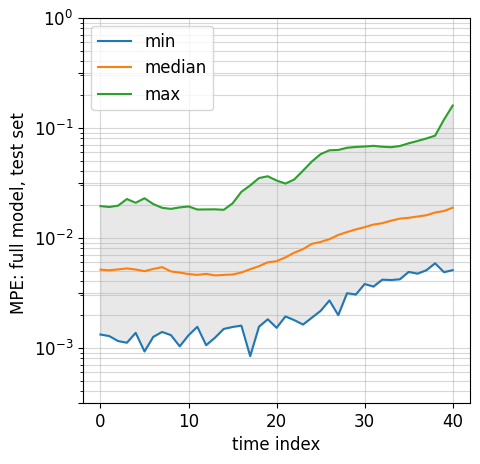} 
\end{tabular}
\end{center}
\caption{Performance of F-ROMs in modeling RMI over all profiles in Table \ref{tab:initialcoeffs} according to the minimum, median, and maximum JL (top) and MPE (bottom) metrics \eqref{eq:metrics} over the test set as a function of time.}
\label{fig:allcoupling_minmax}
\end{figure}

Lastly, we consider modeling all 32 profiles in Table \ref{tab:initialcoeffs} and dispense with the $n_z=2$ model, which previous experiments have shown to be inadequate. Results are shown in Figures \ref{fig:allcoupling} and \ref{fig:allcoupling_minmax}. In this case, the gap closes slightly between $n_z=3$ and $n_z=4$, with $n_z=4$ leading to a more accurate V-ROM, but again failing to generalize as well as the $n_z=3$ F-ROM. This case concludes that the three-dimensional linear dynamics representation of the RMI can successfully model the spectrum of interface perturbations considered over the given EOS parameters and initial velocities. 

\section{Discussion}\label{sec:discussion}

We have demonstrated that the RMI in convergent geometries, parametrized over interface harmonics, initial velocity, and EOS parameters, admits a representation as a three-dimensional linear dynamical system into the nonlinear growth regime. The median MPE of this representation over the test set remains below $2\%$ in each coupling regime, over the time domain considered (Figs. \ref{fig:weakcoupling_minmax}, \ref{fig:strongcoupling_minmax}, \ref{fig:allcoupling_minmax}). The fact that a single nonlinear mapping (given by the trained autoencoder) is capable of compressing all such simulations, across different initial interfaces, is surprising. That the resulting latent representation can be effectively connected back to system parameters is encouraging for future applications. 

The representation can be successfully identified using nonlinear dimensionality reduction via a shallow autoencoder, with linear dynamics enforced using an equation error loss function, utilizing the LaSDI framework. Where previous applications of LaSDI utilized full-field solution measurements, the present use-case models a material bitmap, a vastly undersampled observable of the full dynamics, but one that has been shown in previous studies to be accessible from radiographic images using attention-based neural networks \cite{serino2024learning,serino2024reconstructing}. In future work we aim to combine these techniques with the present methodology to improve inference and control capabilities of hydrodynamic experiments.

Our modeling framework includes a nonlinear mapping from parameters to latent space coefficients and initial conditions, demonstrating that the solution space can be interpolated to draw predictions from new parameter values. Parameters-to-coefficients maps utilized in previous LaSDI studies included predominantly bivariate splines, RBFs, and Gaussian processes. We include in \ref{app:coeffs} a description of some of the coefficient distributions in the current study, arguing that challenges to interpolation in the present use-case fall outside of the scope of previous methods, but are adequately handled using a neural network model for $\CalM$. In particular, Figure \ref{fig:coeff_distrib} demonstrates the multi-modality of individual coefficient distributions, rendering traditional interpolation methods inefficient. We note that latent space interpolation is still an active area, for instance it may be possible to sample thermodynamically consistent EOS parameters as explored in \cite{kevrekidis2024neural}. We aim to incorporate these and other improvements in future work. 

The framework we have introduced can certainly be improved. Figure \ref{fig:latentspace} in \ref{sec:latentspace_vis} includes visualizations of the latent space for $n_z=3$, showing that simulations effectively form clusters that can be associated with their dominant interface modes. In future work we aim to directly incorporate this information into the parameters-to-coefficients map. Additional fine tuning of hyperparameters, including loss coefficients, early stopping regiments, parameters-to-coefficient architectures, etc, can be expected to improve performance. We can conclude that $n_z=2$ leads to inefficient models (indeed this can be visualized in the latent space as the inability to avoid trajectory intersections), however it may be possible to improve the $n_z=4$ model with further tuning. Nevertheless, the success of a linear dynamics representation with $n_z=3$ warrants further study. We strive to offer theoretical evidence of this drastic dimensionality reduction in future work. 

\section*{Acknowledgments}
  The authors would like to thank Youngsoo Choi for helpful suggestions and conversations. This work was supported by the Laboratory Directed Research and Development program at Los Alamos National Laboratory (LANL), under contract No. 20240858PRD1.  


\bibliographystyle{elsarticle-num} %
\bibliography{refs.bib}

\clearpage

\appendix
\addappheadtotoc

\section{Connection between observables and full-order dynamics }\label{sec:advection}
Here we present a brief proof that material bitmaps (the observables considered here) themselves are advective solutions depending on the (unobserved) velocity field. The following is corroborated with much less detail in \cite{herrmann2008nonlinear}, eq. 3.8, and extensively in \cite{osher1988fronts} eq 3.15, for smooth level-set functions, whereas bitmaps are discontinuous. 

Consider a moving set $V(t)\subset \Rbb^d$ and quantity $\varphi(x,t) : \Rbb^d\times \Rbb\to \Rbb$. Then it holds that 
\[\frac{d}{dt} \int_{V(t)} \varphi(x,t)dx = \int_{V(t)}\partial_t\varphi(x,t) dx + \int_{\partial V(t)} \varphi(x,t) \widehat{\nbf}(x,t)\cdot \vbf(x,t)dS(x) \]
\[= \int_{V(t)}\left(\partial_t\varphi(x,t) +\nabla\cdot (\varphi(x,t) \vbf(x,t))\right)dx  \]
where $\widehat{\nbf}$ is the unit normal to the boundary of $V(t)$ and $\vbf(x,t)$ is the velocity at each point in $x\in V(t)$. Now consider the bitmap $U(x,t) = \ind{\rho(x,t)>0}(x)$ where $\partial_t\rho + \nabla \cdot(\rho\ubf) = 0$. Using the formula above, we have
\begin{align*}
\frac{d}{dt} \int_{\Rbb^d}\varphi(x,t)U(x,t)dx &=
\frac{d}{dt} \int_{\rho>0}\varphi(x,t)dx \\
&= \int_{\rho>0}\left(\partial_t\varphi(x,t) + \nabla \cdot (\varphi(x,t) \ubf(x,t))\right) dx \\
&= \int_{\Rbb^d}\left(\partial_t\varphi(x,t) + \nabla \cdot (\varphi(x,t) \ubf(x,t))\right) U(x,t) dx
\end{align*}
Supposing now that $\varphi$ is compactly supported in time, integrating in time we get that
\[\int_0^\infty\int_{\Rbb^d}\left(\partial_t\varphi(x,t) + \nabla \cdot (\varphi(x,t) \ubf(x,t))\right) U(x,t) dxdt = 0\]
so that $U$ weakly solves the advection equation
\[\partial_tU + \ubf \cdot \nabla U = 0\]

\section{Latent space visualization}\label{sec:latentspace_vis}

\begin{figure}
\begin{tabular}{@{}l@{}l@{}l@{}}
    \includegraphics[trim={400 100 350 100},clip,width=0.3\textwidth]{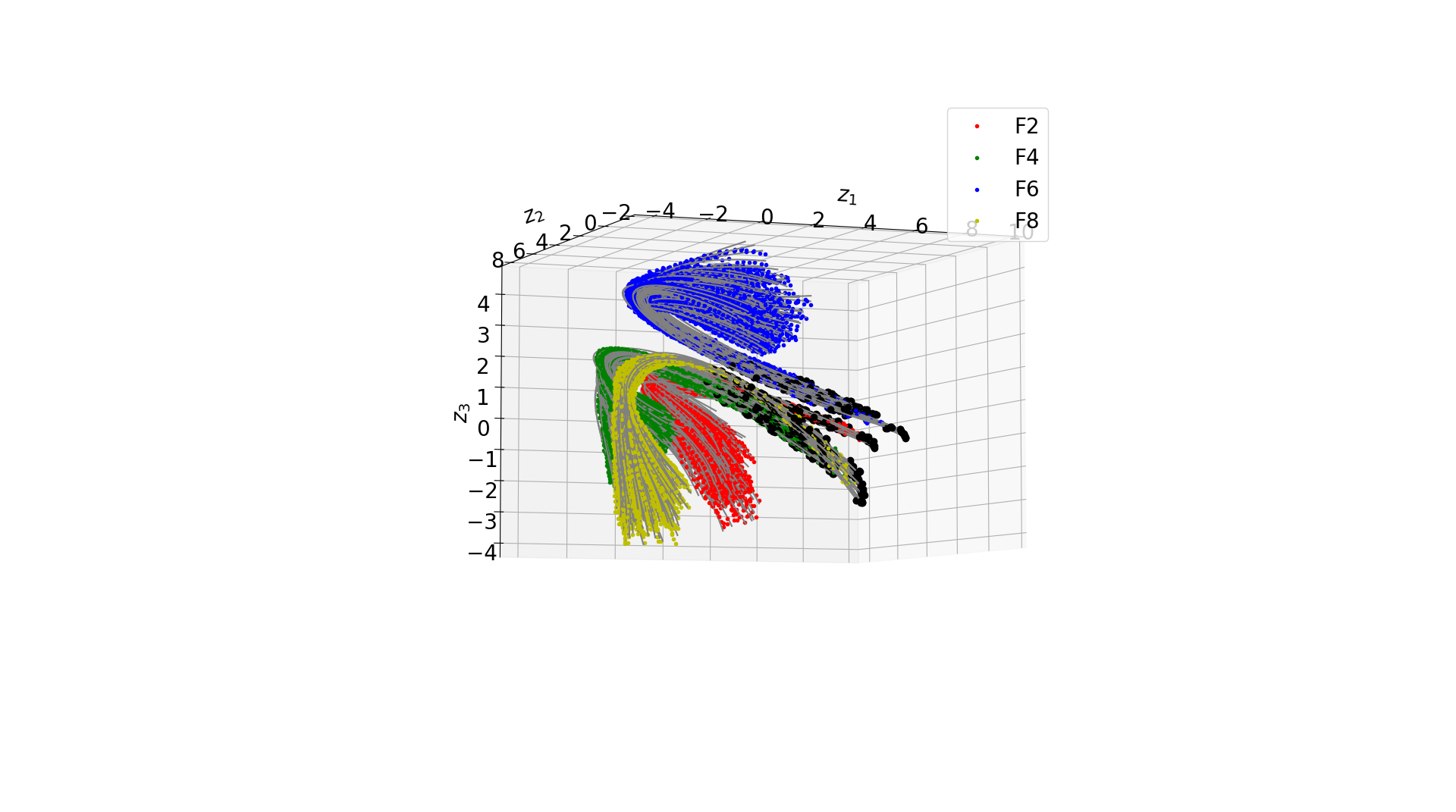} &
    \includegraphics[trim={800 300 750 150},clip,width=0.33\textwidth]{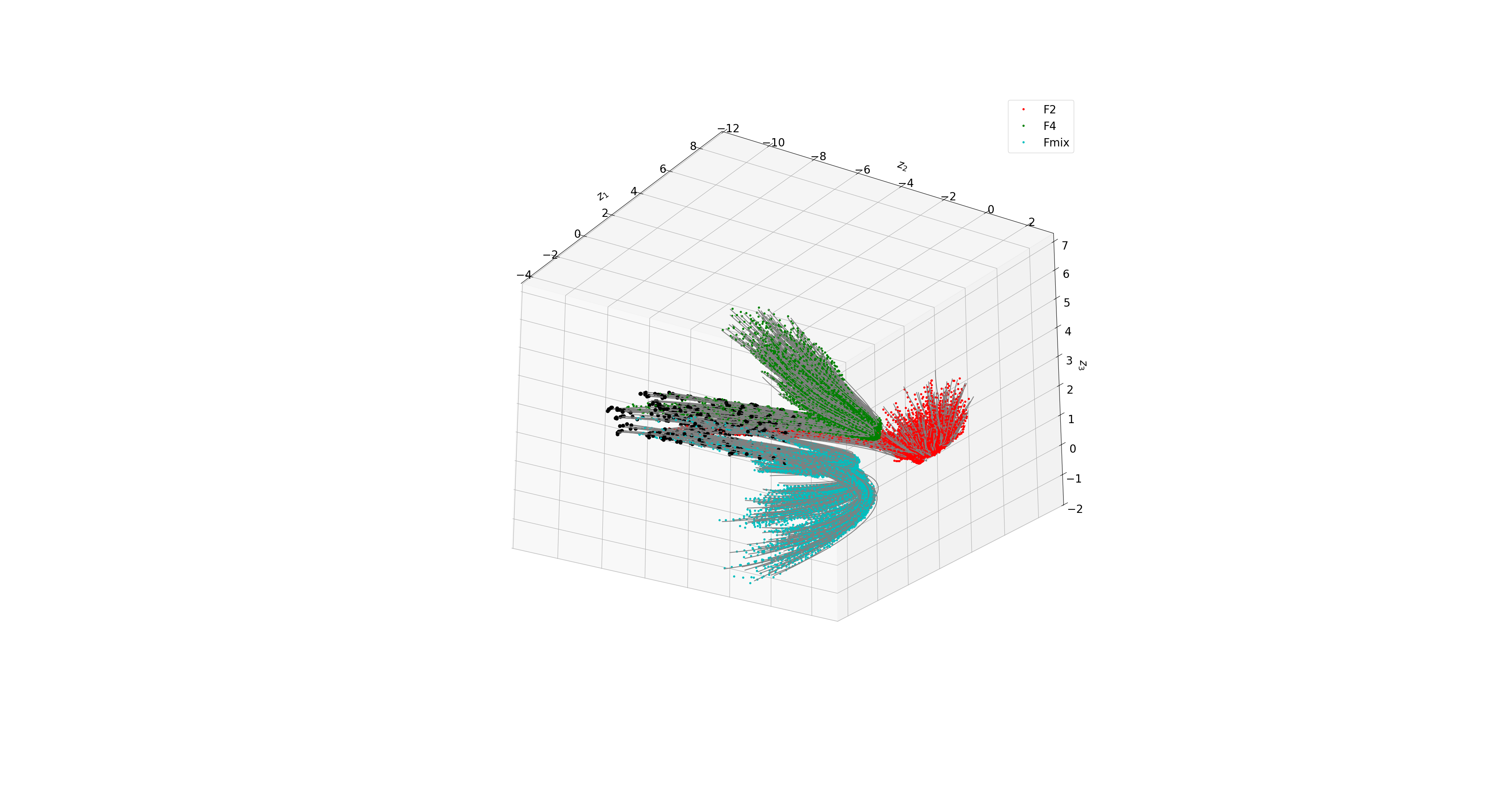} &
    \includegraphics[trim={400 100 350 100},clip,width=0.3\textwidth]{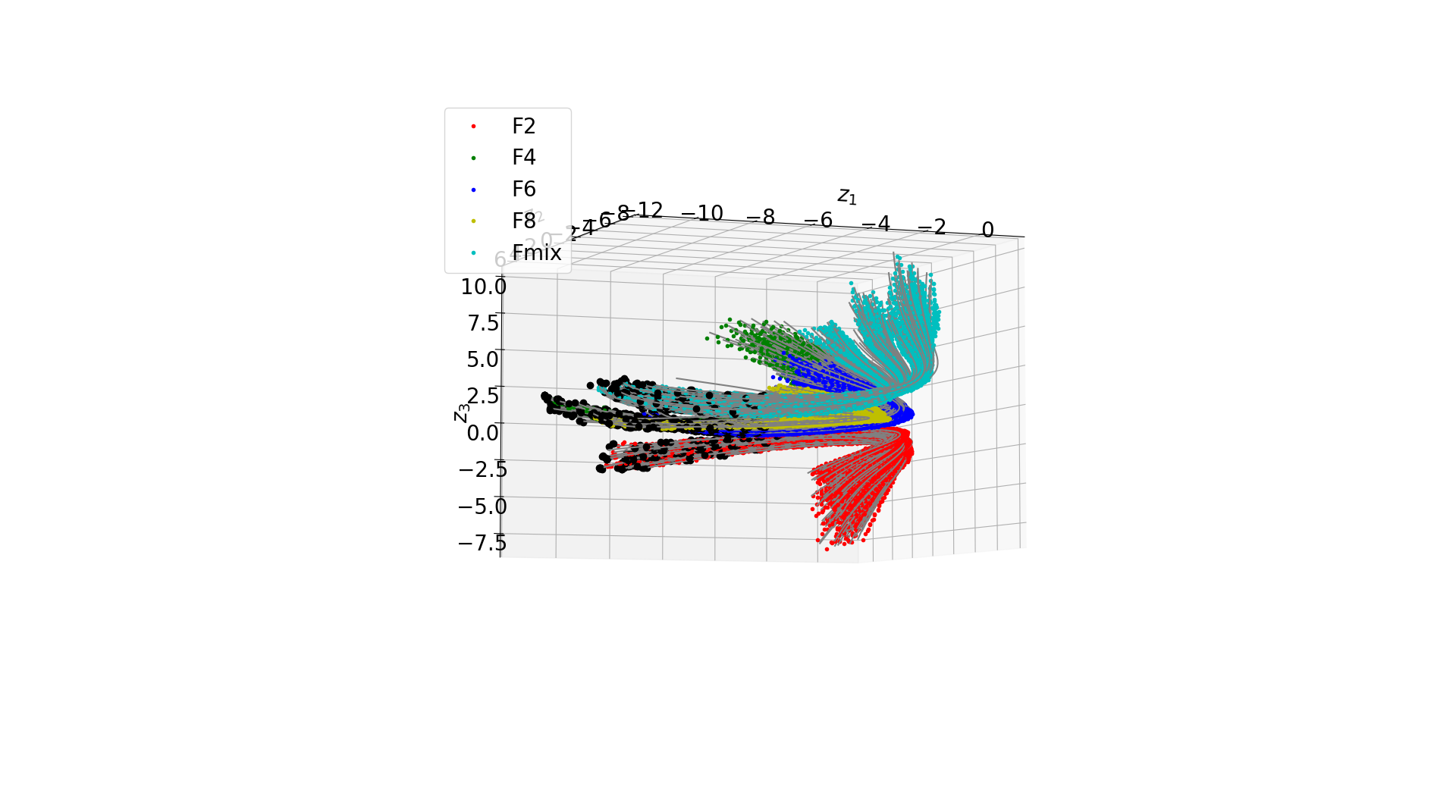}
\end{tabular}
\caption{Visualizations of the latent space trajectories in ROMs $n_z=3$ latent space dimensions for (left to right) weak-coupling, strong coupling, and all profiles. Black dots represent initial conditions. The ROM outputs are plotted in gray over the observed data passed into the latent space via the encoder, which has been colored according to the dominant modes present.}
\label{fig:latentspace}
\end{figure}

In Figure \ref{fig:latentspace} we visualize the latent space in the $n_z=3$ models for the three experiments above. Trajectories are colored according to their dominant interface mode, with strongly mixed modes plotted in cyan. In each case trajectories are found to lie on a disconnected manifold with individual clusters corresponding to dominant modes. The dynamics of all trajectories are similar during the onset of RMI, followed by visible divergence of trajectories as the RMI progresses, leading to distinct mode-separated clusters at later times. In the second and third plots, it can be seen that profiles with strong mixing (cyan) initially form a single cluster, which splits into three sub-clusters during later time evolution corresponding to the three initial mixing strengths in profiles 27-32 (Table \ref{tab:initialcoeffs}). 

\section{Coefficient distributions}\label{app:coeffs}

\begin{figure}
\begin{tabular}{@{}c@{}c@{}}
    \includegraphics[trim={0 0 0 0},clip,width=0.5\textwidth]{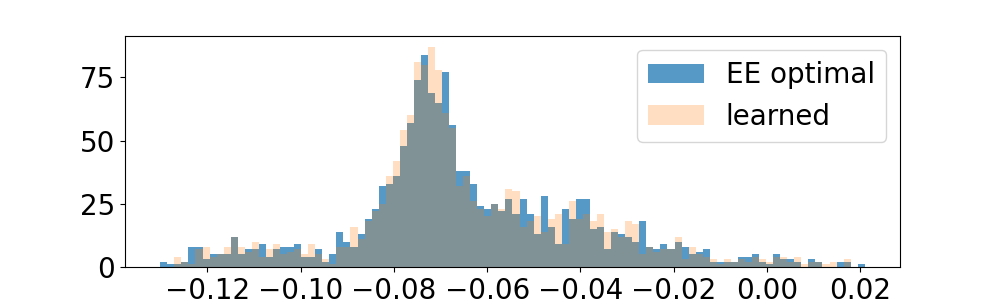} &  
    \includegraphics[trim={0 0 0 0},clip,width=0.5\textwidth]{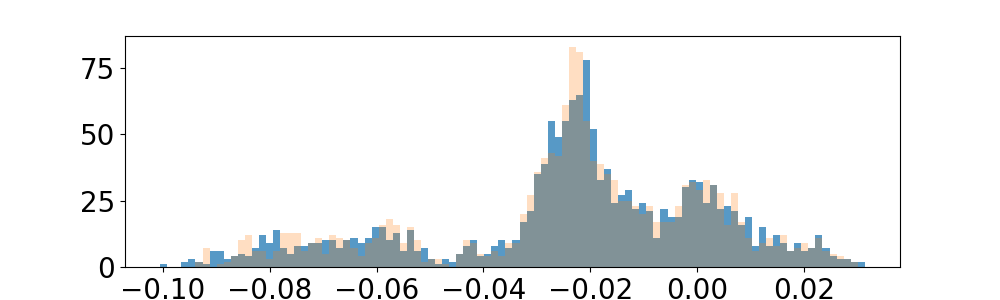} \\  
    \includegraphics[trim={0 0 0 0},clip,width=0.5\textwidth]{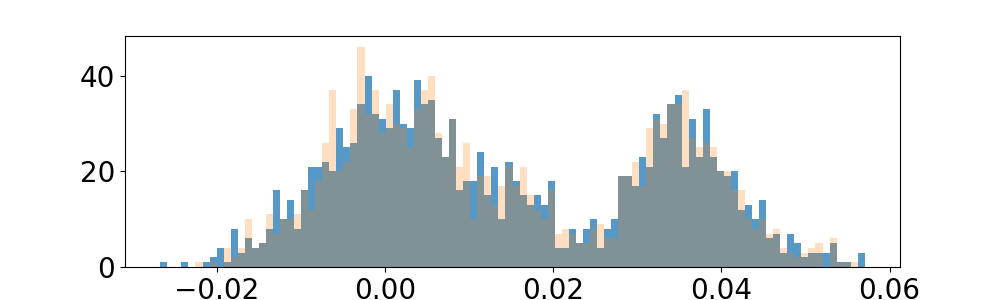} &  
    \includegraphics[trim={0 0 0 0},clip,width=0.5\textwidth]{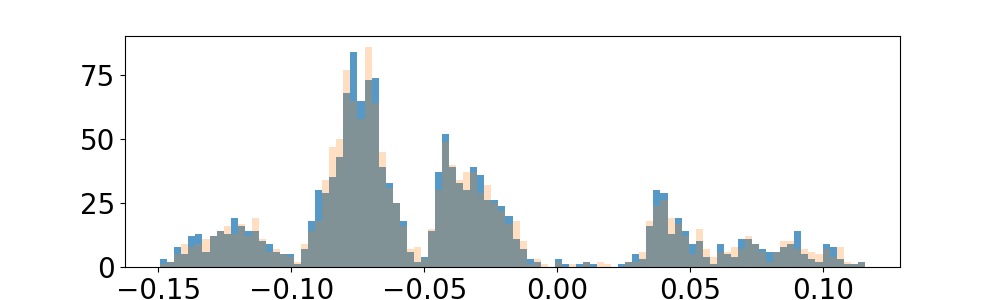} 
\end{tabular}
\caption{Individual distributions of first four ODE coefficients $\wbf$, from the $n_z=3$ model of all profiles in Table \ref{tab:initialcoeffs}.}
\label{fig:coeff_distrib}
\end{figure}

In Figure \ref{fig:coeff_distrib} we plot histograms of four out of the nine coefficients $\wbf$ from the $n_z=3$ F-ROM, comparing the equation-error optimal (EE Optimal) values to their corresponding $\CalM$ outputs. EE optimal coefficients refers to the equation-error least-squares optimal coefficients found from embedding the simulation in the latent space, whereas learned coefficients are the corresponding outputs of the parameters-to-coefficients map. From this it can be seen that traditional interpolation methods should not be expected to perform well. The neural-network mapping $\CalM$ on the other hand successfully captures the complex multimodal distribution for each coefficient.

\section{Comparison between POD and random AE weight initialization}\label{app:PODvsrandom}

\begin{figure}
\begin{center}
\begin{tabular}{@{}c@{}c@{}}

    \text{POD} & \text{Random} \\
    \includegraphics[trim={0 6 0 0},clip,width=0.33\textwidth]{figures/3nz_modes2,4,6,8_nomix_iso3JL_full_model__test_set_.png} &

    \includegraphics[trim={0 6 0 0},clip,width=0.33\textwidth]{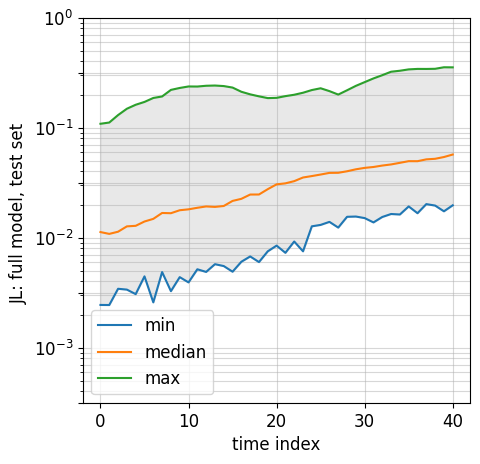} \\
    
    \includegraphics[trim={0 6 0 0},clip,width=0.33\textwidth]{figures/3nz_modes2,4,6,8_nomix_iso3MPE_full_model__test_set_.png} &

    \includegraphics[trim={0 6 0 0},clip,width=0.33\textwidth]{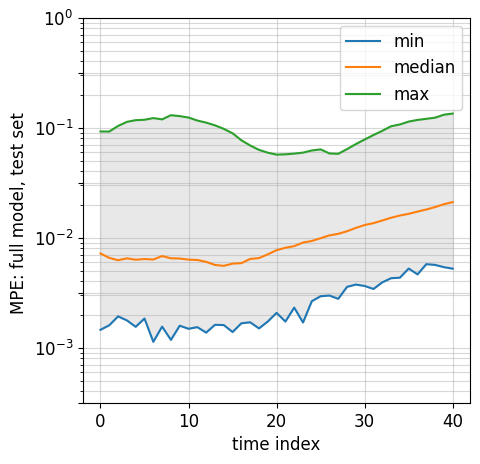} 
    
\end{tabular}
\end{center}
\caption{Comparison between F-ROMs initialized at POD modes (left) and at random weights (right) in modeling RMI in the weak coupling regime (profiles 1-26 in Table \ref{tab:initialcoeffs}) with latent space dimension $n_z=3$.}
\label{fig:randinit}
\end{figure}

Here we show evidence that POD initialization, together with the added regularization loss $\CalR_3$, decreases training time and leads to more accurate ROMs. For brevity we include only results for the weak coupling case (profiles 1-26 in Table \ref{tab:initialcoeffs}) and $n_z=3$, although the results are observed in general. As above, we train three models each with the architecture in Figure \ref{fig:networkdiagram} but with random AE weight initialization. Figure \ref{fig:randinit} shows results from the best performing network (right) in comparison with the analogous results from a POD-initialized AE (left, reproduced from Fig. \ref{fig:weakcoupling_minmax}). Performance on the test set includes a much higher occurrence of outliers and poorer generalization, with maximum errors often more than double that of a POD initialized AE. In addition, both training stages take approximately 20\%-50\% more epochs before the early stopping criterion is satisfied. From this we argue that connection between nonlinear and linear compression schemes is worthy of further study.  

\end{document}

%% file: preamble.tex
\usepackage[utf8]{inputenc}
\usepackage{
amsmath,
amssymb,
amsthm,
appendix,
bm,
bbm,
caption,
color,
epsf,
enumitem,
float,
graphicx,
hyperref,
listings,
mathtools,
mathrsfs,
subcaption,
tikz,
titletoc,
url,
ulem,
xcolor
}

\usepackage[margin=1in,top=1.25cm]{geometry}
\hypersetup{
    colorlinks=true, 
    linktoc=all,     
    linkcolor=black,  
}

\numberwithin{equation}{section}

\tikzset{every label/.style={font=\footnotesize,inner sep=1pt}}
\DeclareCaptionFormat{myformat}{#1#2#3\hrulefill}
\graphicspath{{figures/}}

%% file: notation.tex
\newcommand{\Rbb}{\mathbb{R}}

\newcommand{\Zbb}{\mathbb{Z}}

\newcommand{\Abf}{\mathbf{A}}

\newcommand{\Dbf}{\mathbf{D}}

\newcommand{\fbf}{\mathbf{f}}
\newcommand{\Fbf}{\mathbf{F}}

\newcommand{\Ibf}{\mathbf{I}}

\newcommand{\kbf}{\mathbf{k}}

\newcommand{\nbf}{\mathbf{n}}

\newcommand{\Obf}{\mathbf{O}}

\newcommand{\ubf}{\mathbf{u}}
\newcommand{\Ubf}{\mathbf{U}}
\newcommand{\vbf}{\mathbf{v}}

\newcommand{\wbf}{\mathbf{w}}
\newcommand{\Wbf}{\mathbf{W}}

\newcommand{\Zbf}{\mathbf{Z}}

\newcommand{\vep}{\varepsilon}

\newcommand{\CalF}{{\mathcal{F}}}

\newcommand{\CalL}{{\mathcal{L}}}
\newcommand{\CalM}{{\mathcal{M}}}

\newcommand{\CalR}{{\mathcal{R}}}

\newcommand{\ind}[1]{\mathbbm{1}_{#1}}

